\documentclass[journal, a4paper]{IEEEtran}
\usepackage[T1]{fontenc}
\usepackage[latin9]{inputenc}
\usepackage{color}
\usepackage{amsmath}
\usepackage{amsthm}
\usepackage{amssymb}
\usepackage{esint}
\usepackage{verbatim}
\usepackage{graphicx}
\usepackage{stfloats}
\usepackage{subfigure}
\usepackage{dsfont}
\usepackage{bm}
\usepackage{multicol}
\usepackage{multirow}
\usepackage{tabularray}
\usepackage{makecell}
\usepackage{algorithm}
\usepackage{algorithmic}
\usepackage{mathtools}
\usepackage{array}
\usepackage{xcolor, soul}

\usepackage[numbers,sort&compress]{natbib}

\makeatletter

\usepackage[normalem]{ulem}

\usepackage[colorlinks=true,
linkcolor=blue,
citecolor=blue,
urlcolor=blue]{hyperref}

\newtheorem{assp}{Assumption}

\definecolor{sblue}{RGB}{0,0,0}
\newcommand\re[1]{\textcolor{sblue}{#1}}
\begin{document}
	\title{ 
		  \huge{LiTCom: A Lightweight Transmitter and Inference-Capable Receiver Framework for 6G Uplink}
	}
		\author{Chunmei Xu,~\IEEEmembership{Member,~IEEE},  Siqi Zhang, Zhi Ding, ~\IEEEmembership{Fellow,~IEEE}, \\ Yi Ma,~\IEEEmembership{Senior Member,~IEEE},  Rahim Tafazolli,~\IEEEmembership{Fellow,~IEEE} 
			
			
			\thanks{C.~Xu, S. Zhang, Y. Ma and, R. Tafazolli are with 5GIC \& 6GIC,
				Institute for Communication Systems (ICS), University of Surrey, Guildford,
				U.K. (emails:\{chunmei.xu; s.zhang; y.ma; r.tafazolli\}@surrey.ac.uk). 
			}
		 
			\thanks{Z. Ding is with the Department of Electrical and Computer Engineering, University of California at Davis, Davis, CA 95616 USA (e-mail: zding@ucdavis.edu).}
		}

		\maketitle
		\begin{abstract}
			
		This paper introduces LiTCom, a lightweight transmitter and inference-capable receiver framework, designed to enable robust 6G uplink communication under low signal-to-noise (SNR) conditions. 
		It embraces the resource asymmetry between edge devices and the network infrastructure.   
		LiTCom simplifies transmitter design by applying basic low-pass filtering for source coding and minimal channel coding, significantly reducing the processing complexity. 
		The receiver employs large-scale generative artificial intelligence (GenAI) models to infer high semantic-fidelity content from highly distorted and degraded signals beyond traditional decoding capabilities. 
		Furthermore, efficient power allocation strategies are developed by exploiting data importance to improve system performance, which is measured by the introduced quality of experience (QoE) metric. 
		Simulation results validate the effectiveness of the proposed LiTCom framework and the lightweight coding design. \re{Compared with the 5G NR-like baseline (using JPEG source coding and LDPC channel coding) and the Deep-JSCC baseline,  LiTCom achieves SNR gains up to $8$ dB and $2.5$ dB,} respectively, while reducing over 95\%  transmitter-side  computations.

	  
		\end{abstract}
		
		
		\section{Introduction}	
		Next generation wireless networks (6G) are envisioned to support emerging intelligent applications such as immersive extended reality (XR), telepresence,  digital twins, and autonomous robots \cite{giordani2020toward, wang2023road}. These applications require continuous acquisition of high-dimensional sensory data and real-time interpretation of the physical environment, generating persistently heavy data traffic. This imposes stringent requirements on the wireless network in terms of throughput and latency.  Moreover, these applications depend more on the task-relevant semantics, such as scene understanding, facial expression, or object positions, and highlights quality of experience (QoE) of users,  relaxing the requirement on bit-exact reconstruction \cite{yang2022semantic}.

		 Importantly, the sensory data, such as  images, videos, and contextual observations, originate at the edge devices, making the uplink the primary carrier of  intelligence-critical data.  However,  edge devices, such as wearables or mobile terminals, are often severely constrained in resources of computation, processing energy, transmit power, and bandwidth.  
		Such computation and energy limitations hinder high-complexity processing on the source data, resulting in degraded performance and increased  processing latency. In addition, their scarcity of radio resource further impede high-throughput and low-latency transmissions, causing limited coverage and frequent retransmissions. 
		In contrast, the network infrastructure such as base stations (BS)  benefits from substantially more powerful computation platforms,  higher transmit power, wider bandwidth, and larger antenna arrays. Moreover, the downlink typically demands far lighter control or feedback information than the uplink, since the network acts as the  decision maker and resource scheduler \cite{3gpp_ts_38_300}. 
		Therefore, \re{the uplink, rather than the downlink,  becomes} the principal bottleneck for enabling these emerging intelligent 6G applications \cite{AIRAN2026AIonRAN}. This motivates our focus on the uplink in this work.

		\re{Supporting these applications in the uplink  requires a fundamental rethinking on two  aspects of communications: 1) how the source data is processed; and 2) how the processed data is delivered.}
	 	The traditional bit-centric communication paradigm is primarily designed to optimize the quality of service (QoS) in terms of data rate, latency, and reliability rather than the QoE. In processing source data, traditional coding schemes aim to minimize representation length by exploiting inherent data characteristics to remove redundancy, subject to an acceptable distortion constraint  \cite{wallace1992jpeg, wiegand2003overview}. Such schemes emphasize bit-level fidelity and overlook semantic relevance with downstream tasks, {which may produce higher data traffic due to the transmission of irrelevant data \cite{gunduz2022beyond}}. Moreover, producing highly compact representation of multi-dimensional source requires substantial processing complexity, which can be unaffordable for resource-constrained edge devices.
	 
	 	The representing compactness also introduces strong inter-symbol dependencies, making the reconstruction highly vulnerable to errors. Thereby, bit-perfect transmission of the processed information under wireless impairments is necessitated in the traditional paradigm. To combat channel impairments, sophisticated channel coding schemes have been developed and advanced for error correction \cite{hamming1950error, viterbi1971convolutional, berrou1993near, richardson2001design}.  However, due to the limited transmit power of the edge devices, the uplink transmissions are more likely to operate in low-SNR conditions. Existing channel codes degrade significantly, known as the cliff effect, and may even underperform channel-uncoded transmission. In addition, the limited bandwidth and small antenna apertures of the edge devices impose a strict ceiling on the achievable channel capacity. Consequently, traditional  bit-centric communication systems are increasingly incapable of meeting the stringent requirements of 6G applications that depend on real-time, context-aware understanding of the physical world.  
	   
		Semantic communication (SemCom) has emerged as a promising paradigm beyond the conventional bit-centric one \cite{gunduz2022beyond, kountouris2021semantics}. \re{Substantial efforts have been devoted to developing theoretical foundations for SemCom based on semantic entropy, knowledge graphs, rate-distortion-perception theory, and the information bottleneck principle. However, a unified theory that can  guide system design and implementation remains underdeveloped \cite{shao2024theory}. As a result, most existing works rely on neural network (NN)-based methods to extract task-relevant semantic information from source data at the transmitter and recover or infer the intended content at the receiver.} 
		\re{A representative category is to adopt end-to-end NN architectures, including deep joint source-channel coding (Deep-JSCC), which jointly learn the semantic encoder and decoder and transmit encoded data in the analog manner  \cite{xie2021deep, weng2021semantic, bourtsoulatze2019deep, xu2023JSCC, erdemir2023generative, liu2025resitok}. In order to improve adaptability, flexibility and compatibility in practical deployment, extensions have further incorporated digital modulation and orthogonal frequency division multiplexing schemes \cite{liu2024ofdm, zou2025analog}, as well as incremental redundancy hybrid automatic repeat request  mechanisms \cite{jiang2022wireless}.}

		\re{Despite performance gains over conventional bit-centric systems, these end-to-end NN models are often trained under specific datasets, channel models, compression rates, and SNR regimes. Their generalization to scenarios that deviate from the training conditions, is therefore limited. In addition, the evolution towards networked SemCom with multiple users introduces new challenges, including increasing training complexity, knowledge-base alignment across distributed devices, and dynamic network topology \cite{guo2024survey}. Beyond such learning-based SemCom, recent studies have investigated generative SemCom that directly deploys pre-trained models at both ends \cite{xu2025generative, qiao2024latency}, where semantic importance can be modelled and exploited for adaptive transmission \cite{xu2025dataimportanceJ}. This framework can improve generalization in diverse settings and ensuring compatibility with existing digital systems.  However, both the learning-based SemCom and the generative SemCom frameworks require substantial transmitter-side computation for  neural encoding and semantic extraction. This computational burden can exceed the capability of severely resource-constrained edge devices, thereby limiting their practicality for 6G uplink scenarios.}

		Therefore, a new communication paradigm is needed for 6G uplink, which enables computationally lightweight processing at the transmitter while maintaining robust information delivery under low-SNR regions.  \re{This paradigm is facilitated by emerging AI-RAN architectural trends toward cloud-native, accelerated edge infrastructure \cite{11320975}, together the recent advances in generative artificial intelligence (GenAI).}  In particular, large-scale diffusion-based models have achieved remarkable breakthroughs, which are trained on diverse and multimodal datasets spanning from text, audio, images, and videos  \cite{chen2024big,10384630}. These GenAI models exhibit strong generative priors, which can be leveraged to infer missing details and produce high-quality reconstructions from incomplete or degraded inputs \cite{radford2021learning, brown2020language, rombach2022high,zhang2023adding, yu2024scaling}.
	  	When deployed as a generative source decoder, the receiver can potentially infer QoE-satisfying reconstruction from the representations that may be distorted during compression and degraded through wireless transmission.  By shifting the burden of reconstruction from bit-exact recovery to semantic inference,  the requirement for strict bit-compact compression and bit-perfect transmission can be significantly relaxed.  \re{This motivates LiTCom, a lightweight transmitter and inference-capable receiver framework, for 6G uplink.}

		The main contributions of this work are summarized:
		

	 \begin{itemize}
	  		\item We propose a novel {inference-oriented communication} framework for 6G uplinks, featuring a lightweight transmitter and an inference-capable receiver. The proposed lightweight coding design is well suited for the computation- and energy-constrained edge devices. The GenAI models empower the receiver to infer QoE-satisfying content from distorted and channel error-degraded representations, allowing edge devices to transmit robustly across low-SNR regimes. 
	  		
			\item We establish the theoretical foundations of LiTCom by characterizing two fundamental properties of representing sufficiency and error robustness. \re{Guided by these, the design principle of the lightweight semantic-preserving source coding and error-distributing channel coding are provided for images}.
			
			\item \re{We define the perceived uplink coverage based on the proposed QoE measure} that combines the natural image quality evaluator (NIQE) and contrastive language-image pre-training (CLIP)-based similarity. To improve the system performance,  efficient importance-aware power allocation strategies are developed with closed-form waterfilling solutions.
			
			\item Simulation results validate the effectiveness of the proposed LiTCom framework and its lightweight coding design. Compared with conventional \re{5G NR-like} and Deep-JSCC baselines, LiTCom achieves up to 8 dB and 2.5 dB SNR gains, respectively, while reducing over 95\%  transmitter-side  computations. 			  
		\end{itemize}

		\section{ Principles of LiTCom   \label{sec:II}}
		LiTCom is designed to enable a resource-constrained transmitter to perform robust 6G uplink transmission under low-SNR conditions, prioritizing semantic QoE over traditional QoS performance.
		In a simplified mathematical form, a wireless communication process can be modelled as: 
		 \begin{equation}
		 	\mathbf{I}\overset{(a)}{\rightarrow} \mathbf{s} \overset{(b)}{\rightarrow} \hat{\mathbf{s}}  \overset{(c)}{\rightarrow} \hat{\mathbf{I}}_{\hat{\mathbf{s}}},
		 \end{equation}where $\mathbf{I}$,  $\mathbf{s}$, $\hat{\mathbf{s}}$, and  $\hat{\mathbf{I}}_\mathbf{s}$ denote the original source, the transmitted representation,  the received representation, and  the reconstructed source, respectively. Subprocesses $(a)$ corresponds to source processing, $(b)$ represents the operation in between, including channel coding, modulation, wireless transmission, channel decoding and demodulation, etc.,  and $(c)$ denotes source decoding.  We focus on the perceptual and semantic quality of the reconstruction   $\hat{\mathbf{I}}_{\hat{\mathbf{s}}}$ in this paper. Such quality is assessed using the semantic QoE metric (detailed in Sec. \ref{sec:IV}), with lower semantic QoE values indicating better performance.

		 
		 \subsection{Theoretical Framework}		
		   \re{Under LiTCom framework, the resource-rich receiver deploys the large-scale GenAI model as the generative source decoder. The inference process can be expressed by:}    
			\begin{equation}\label{eq05}
				\hat{\mathbf{I}}_{\hat{\mathbf{s}}} = \mathcal{G}_\theta(\hat{\mathbf{s}}),
			\end{equation}
			where $\mathcal{G}_\theta(\cdot)$ denotes the pre-trained GenAI model. The model parameters $\theta$ capture the generative prior, enabling the receiver to infer semantic QoE-satisfying reconstructions from incomplete inputs \( \hat{\mathbf{s}} \). In a communication system,  such incompleteness arises from compression distortion in subprocess $(a)$ and/or transmission degradation in subprocess $(b)$.  In the sequel, we aim to characterize conditions under which this incompleteness still guarantees the semantic QoE performance.
	
			\subsubsection{Representing Sufficiency}
			A source representation $\mathbf{s}$ is considered to be sufficient when the corresponding reconstruction $\hat{\mathbf{I}}_{\mathbf{s}}$ meets the semantic QoE requirement. We first define the semantic QoE-essential representation $\mathbf{s}^*$,  which is sufficient and necessary such that any additional distortion could lead to an unacceptable semantic QoE. Mathematically,  the QoE-essential representation can be formed by:
				{\begin{equation}
						 \mathcal S^* \triangleq \left\{ 
						\mathbf{s}^* \,\middle|\,
						\begin{aligned}
							& \mathrm{QoE}(\hat{\mathbf{I}}_{\mathbf{s}^*};\mathbf{I}) \le \mathrm{QoE}_\mathrm{th} \\
							& \mathrm{QoE}(\hat{\mathbf{I}}_{{\mathbf{s}}'}; \mathbf{I}) > \mathrm{QoE}_\mathrm{th} 
						\end{aligned}
						\right\},
				\end{equation}}where ${\mathbf{s}'}$ is any distorted version of $\mathbf{s}^*$,
				$\hat{\mathbf{I}}_{\mathbf{s}^*}=\mathcal{G}_\theta(\mathbf{s}^*)$, and $\hat{\mathbf{I}}_{{\mathbf{s}}'}=\mathcal{G}_\theta({\mathbf{s}}')$ are the inferred reconstructions from $\mathbf{s}^*$ and ${\mathbf{s}}'$, respectively.  
			  Function $\mathrm{QoE}(\cdot; \cdot)$ represents the semantic QoE evaluator, and $\mathrm{QoE}_\mathrm{th}$ denotes the satisfactory threshold. 
			  
\begin{figure}[tp]
					\centering
					\includegraphics[width=0.9\columnwidth]{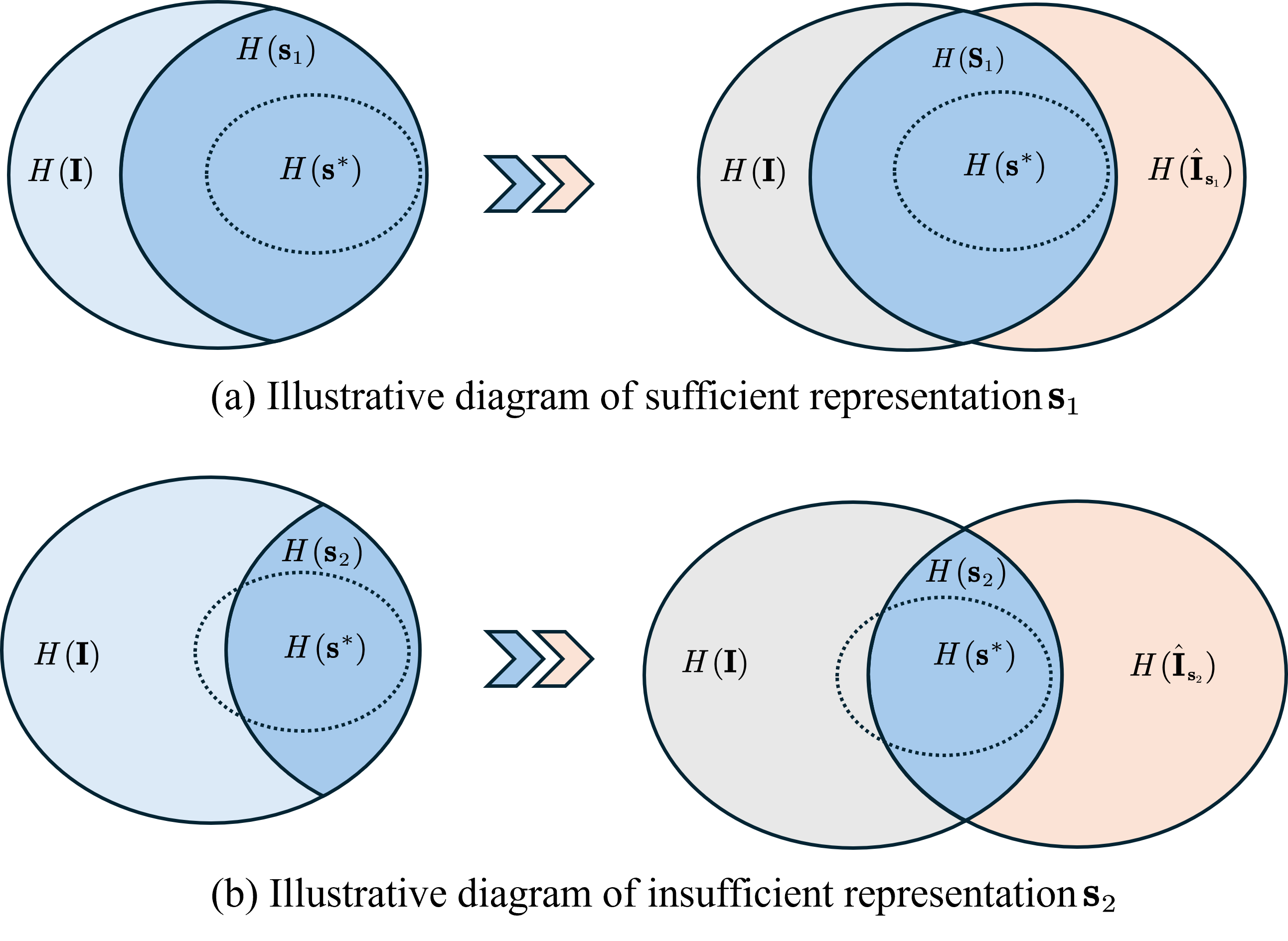}
					\caption{Illustrative diagram of sufficient and insufficient representations, where original regions represent the inferred information at the receiver. $H(\cdot)$ denotes the entropy function. 
					}
					\label{fig:entropy}
				\end{figure}

			 Then, the sufficiency condition of a representation $\mathbf{s}$ can be defined such that it includes a certain QoE-essential representation $\mathbf{s}^*$ in $\mathcal S^*$\re{, which is illustrated in  Fig. \ref{fig:entropy}.} In Fig. \ref{fig:entropy}(a), the semantic QoE is satisfied when $\mathbf{s}^*$ is preserved within $\mathbf{s}_1$. In contrast, reconstruction from insufficient representation $\mathbf{s}_2$ results in unacceptable QoE performance as shown in Fig. \ref{fig:entropy}(b).  The grey regions are the distorted information during compression in subprocess $(a)$, while the orange parts represent the inferred information during subprocess $(c)$. This distorted information \re{typically should be} less critical for semantic fidelity, but often contributes to the perceptual quality.  
				
			\subsubsection{Error Robustness}The received representation $\hat{\mathbf{s}}$ is likely to contain errors caused by channel impairments, particularly under low-SNR conditions, which can be formed as: \begin{equation}\label{eq17}
				\hat{\mathbf{s}} =  \mathbf{s} + \mathbf{e},
			\end{equation}where $\mathbf{e}$ denotes the representation error. It is modelled to follow the distribution  of $ \mathcal D (\boldsymbol{\mu},\boldsymbol{\Sigma})$, where $\boldsymbol{\mu}=\mathbf{0}$ and $\boldsymbol{\Sigma}$ are the mean vector and covariance matrix, respectively. \re{The \re{mean square error (MSE)} is denoted as $\mathrm{MSE}(\mathcal D) = \frac{1}{s} \mathbb E [\Vert\mathbf{e}\Vert^2] = \frac{1}{s}\mathrm{tr}(\boldsymbol{\Sigma})$, where $s$ is the dimension of $\mathbf{e}$, $\mathbb{E}(\cdot)$ stands for the expectation, $\Vert\cdot\Vert$   for the $\ell_2$-norm,  $\mathrm{tr}(\cdot)$ for the trace operator.}
			
			By leveraging the GenAI models, the receiver is capable of semantically compensating for those corruptions by inferring coherent information to ensure the semantic performance.  This is referred to as error robustness of LiTCom, which characterizes the ability to combat channel impairments while ensuring QoE-satisfying reconstructions. Given a sufficient representation $\mathbf{s}$, the error robustness conditions can be formalized by a set of feasible distributions:  
			 \begin{equation} \label{eq6}
			 	\mathcal R^*(\mathbf{s}) \triangleq \left\{\mathcal D^*\middle| \mathbb E_{\mathcal D^*}\big[\mathrm{QoE}(\hat{\mathbf{I}}_{\hat{\mathbf{s}}};\mathbf{I})\big] \le \mathrm{QoE}_\mathrm{th} \text{ given } \mathbf{s}\right\}.
			\end{equation}
			The corresponding set of MSE is  given by:
			 \begin{equation} \label{eq7}
			 	\mathcal E^*(\mathbf{s}) =\left\{\mathrm{MSE}(\mathcal D^*)
			 	\middle|\forall \mathcal D^*\in \mathcal R^*(\mathbf{s})
			 	\right\}, 
			 \end{equation}  
			 

		 To \re{better understand the impact of $\mathcal D^*$ on the semantic performance}, two representative classes of error distributions are considered:
		\begin{itemize}
		 \item[i)] Independent case: The error components are statistically independent, and $\boldsymbol{\Sigma}$ is diagonal. This corresponds to unstructured errors.   Such random perturbations are often semantically insignificant, as the generative prior can implicitly perform denoising, thereby restoring the source with high perceptual quality while maintaining strong semantic similarity with the original one \cite{lin2024diffbir}.  
		
		 \item[ii)] Correlated case: Non-zero off-diagonal elements exist in $\boldsymbol{\Sigma}$, capturing the dependency among error components. This produces structured or sequential error patterns. This type of degradation results in locally coherent blocks of missing information, which can be inpainted by leveraging residual contextual and semantic cues  \cite{yeh2017semantic, yu2018generative}. However, the reconstructed output may not fully preserve semantic consistency with the original one. 
		\end{itemize}

	The above-defined  $\mathcal R^*(\mathbf{s})$ is analytically intractable due to the non-trivial and implicit expressions of both the generative decoder and the semantic QoE evaluator. 
	 Since direct QoE modelling is difficult, QoE approximation using QoS indicators has been extensively studied \cite{QoE2010, zhang2018towards, zhang2023qoe1, zhao2016qoe, zhang2023qoe2, alreshoodi2013survey}. Existing studies have shown that  QoE is generally correlated with QoS indicator in the application-layer, while the QoS indicators across different layers are also positively correlated.  \re{Additionally, the bit error rate (BER), a QoS measure, can be used to approximately derive the MSE of the received representation, generally exhibiting a positive correlation \cite{xu2025dataimportanceJ}. Therefore, we make the following two assumptions for LiTCom:}
 
 \re{
 	\begin{assp}\label{hyp2} 
 		The semantic QoE degradation is positively related to the MSE of the received representation.
 \end{assp}
\begin{assp}\label{hyp1} 
	Under the same MSE level, the semantic QoE degradation of the reconstructed source decreases as the representation errors become more independent and less structured.
	\end{assp}
}

	\re{Under these two assumptions,} the error robustness defined in \eqref{eq6} can be relaxed. Given a specific type of error distribution $\mathbb D$, such as Gaussian distribution,  it can be approximately characterized using the MSE : 
	\begin{equation} \label{eq08}
		\re{\hat{\mathcal R}^*(\mathbf{s}; \mathbb D)  \triangleq \left\{\mathcal D^*\middle|  \mathrm{MSE}(\mathcal D^*) \le \mathrm{ MSE}_{\mathrm{th}} \text{ given } \mathbf{s} \text{ and } \mathbb D\right\}.}
	\end{equation}$\mathrm{ MSE}_{\mathrm{th}}$ denotes the maximum tolerant error, \re{which is jointly determined by $\mathrm{QoE}_\mathrm{th}$, the sufficient representation $\mathbf{s}$, and the error distribution type $\mathbb D$.} 

\subsection{Coding Design Principles}
This subsection discusses the key coding design principles of LiTCom \re{that ensures human perceptual and semantic understanding experience on images}, including the semantic-preserving source coding and error-distributing channel coding schemes. 

\subsubsection{Semantic-Preserving Source Code} \re{The QoE-essential information of image sources is predominantly carried by low-frequency components, such as structures and object layouts. Fine-grained details in  high-frequency components contribute less to semantic understanding and can be plausibly regenerated by powerful GenAI models such as SUPIR through their learned natural-image priors \cite{yu2024scaling}. This motivates a semantic-preserving encoding strategy in which low-frequency components are explicitly preserved, while high-frequency components are selectively discarded  at the transmitter and resynthesised through generative inference at the receiver.}

We propose a lightweight semantic-preserving  source coding scheme based on low-pass filtering (LPF) to preserve the essential low-frequency components.
Specifically, an original visual signal \( \mathbf{I} \in \mathbb{R}^{H\times W} \) is divided into non-overlapping blocks of  size  $B_1\times B_2$.
Applying a low-pass filter (i.e., mean averaging) to each block yields a compressed representation $\mathbf{s}\in\mathbb R^S$. Letting $I=HW$, we have $S=rI$ with $r=1/{B_1B_2}$ being the compression rate. This method requires only $1$ multiplication and $B_1B_2$ additions per block, making it computationally lightweight and suitable for resource-constrained edge devices. The proposed source coding scheme can be expressed as: 
\begin{equation} \label{eq01}
	\mathbf{s} = \mathcal{F}_{\mathrm{sp\text{-}src}}(\mathbf{I}),
\end{equation}producing a low-quality, blurred version of source $\mathbf{I}$.  \re{Inter-symbol} dependency within $\mathbf{s}$ is significantly reduced compared to conventional entropy-based source coders and Deep-JSCC models.  
		
At the receiver, the goal is to infer fine-grained details suppressed during source compression and degraded during wireless transmission. \re{This can be viewed as an image restoration (IR) problem in computer vision.}
State-of-the-art IR techniques have demonstrated remarkable performance in denoising and super-resolution by using generative diffusion models. In particular, the SUPIR model~\cite{yu2024scaling} has shown exceptional capability in ultra-high-quality image reconstruction.  
  The generative source decoder that employs this GenAI model can be written as
\begin{equation}\label{eq05}
	\hat{\mathbf{I}} = \mathcal{F}^{-1}_{\mathrm{sp\text{-}src}}(\hat{\mathbf{s}}) = \mathcal{G}_\theta(\hat{\mathbf{s}}),
\end{equation}where $\mathcal{G}_\theta$ is the mapping of the SUPIR model, and $ \hat{\mathbf{s}} $is the received representation that may be further degraded by channel errors.

\re{Note that SUPIR is adopted as a representative implementation of the generative decoder rather than as a model-specific requirement of LiTCom. Any other generative IR models that can infer QoE-satisfying content from degraded representations can be employed. Such replacements may change the numerical operating point, including the achievable QoE, required SNR, and acceptable error distributions and their tolerant MSE  level, but do not alter the architectural design principles of generative inference.
}

\subsubsection{Error-Distributing Channel Code} 
 
Given the semantic-preserving source coding, the error robustness of LiTCom depends on the residual error distribution. 
\re{Formally, an \textit{error-distributing code} is a channel coding strategy, particularly relevant in low-SNR regimes, that aims to shape the residual error distribution $\mathcal D$ into the perturbation that maximises QoE performance. Such QoE performance maximisation is not necessarily equivalent to the residual bit error minimisation that is the primary objective of conventional forward error correction (FEC) codes.  Nevertheless, FEC can be regarded as a kind of sub-optimal error-distributing codes targeting $\mathcal D$ concentrated near zero, while channel-uncoded transmission  is a special case where $\mathcal D$ tends to be less structured.} 

\re{In terms of the conventional BER metric, strong FEC codes, such as LDPC, Turbo, and Polar codes, typically outperform weak FEC codes, such as convolutional, Hamming, and repetition codes, as well as channel-uncoded transmission, when the SNR is above their effective decoding threshold. However, in very low-SNR regimes, their decoding performance may degrade sharply, and the resulting residual errors can become highly structured and correlated \cite{berrou1993near, richardson2001design, 7998249,gastpar2003code}.} Since LiTCom targets robust operation under low-SNR and transmitter-resource constraints, we adopt weak FEC codes or channel-uncoded transmission as lightweight error-distributing coding schemes in this work. \re{Note that the weak FEC and channel-uncoded choices considered here are representative lightweight designs rather than universally optimal solutions, which are generally dependent on the SNR regime, source representation, and receiver-side generative decoder. Under the considered low-SNR settings, these schemes tend to produce more independent and less correlated residual errors, leading to a smaller MSE than strong FEC codes and making the received representations more amenable to generative reconstruction. When the SNR is sufficiently high, strong LDPC codes can be applied as the error-distributing code.}

The  processes of error-distributing channel encoding and decoding can be formulated as: 
\begin{equation}\label{eq02}
	\mathbf{c} = \mathcal{F}_{\mathrm{ed\text{-}ch}}(\mathbf{s}),  \quad 	\hat{\mathbf{s}} = \mathcal{F}_{\mathrm{ed\text{-}ch}}^{-1}(\hat{\mathbf{c}})
\end{equation}where $\mathbf{c}$ and $\hat{\mathbf{c}}$ are the transmitted and received channel codewords, respectively.  Their relationship can be given by
\begin{equation}
	\hat{\mathbf{c}}=\mathcal{T}(\mathbf{c}; 	\mathrm{snr}),
\end{equation}where $\mathcal{T}$ encapsulates all intermediate operations.  
 
 \re{It is worth noting that the proposed LiTCom framework is most naturally applicable to modalities with exploitable temporal, spatial, or spectral structures and with mature generative restoration models. For example, the same receiver-side inference principle can be extended to audio and video transmission, where the inference-capable receiver may be implemented using BABE~\cite{moliner2024blind} and DiffVSR~\cite{li2025diffvsr}, respectively. By contrast, it is less direct to extend LiTCom to modalities without clear structural redundancy or well-established generative restoration models, such as radar point clouds or haptic signals, and to downstream tasks relying on fine-grained details for precise execution. In such cases, the semantic- or task-relevant representation may need to be learned through NN learning or tokenisation, potentially guided by the IB principle and the structured IB framework~\cite{yang2025structured}. }
	
\section{System Model \label{sec:III}}	
\begin{figure*}[t!]
				\centering
				\includegraphics[width=0.75\textwidth]{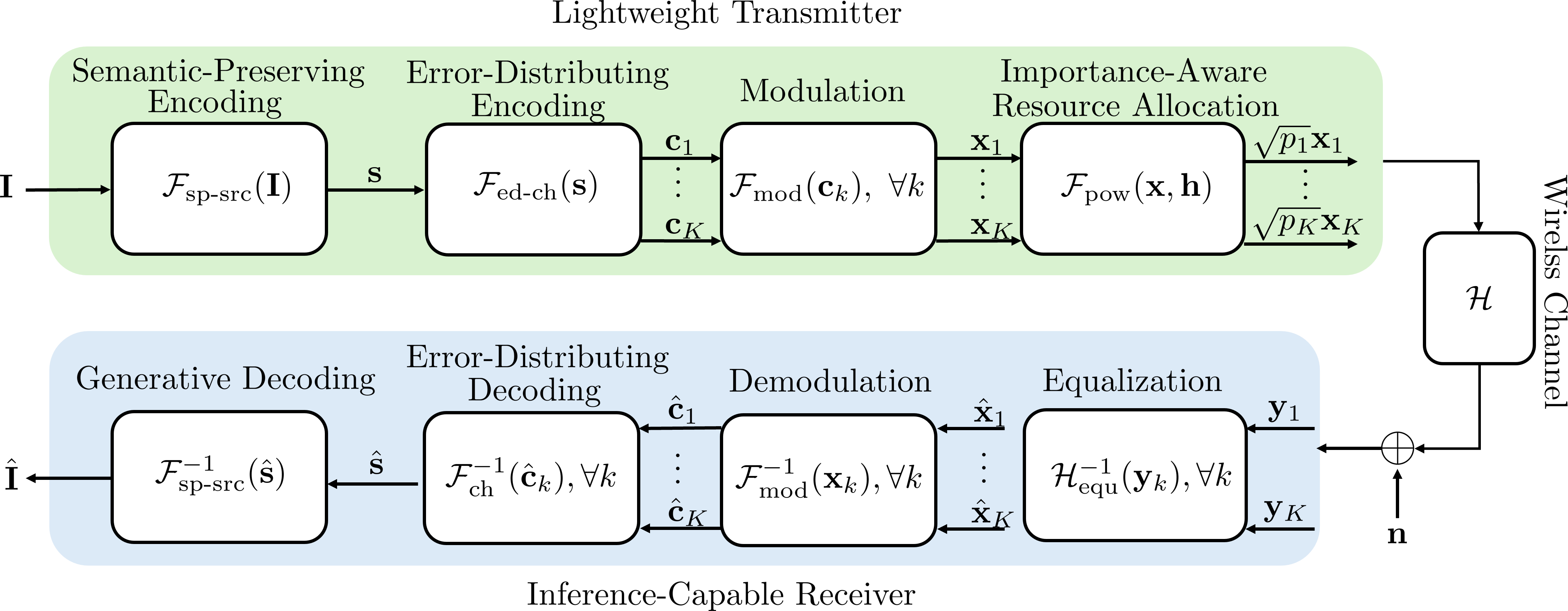} 
				\caption{System model of the proposed LiTCom framework for robust 6G uplink transmission.}\label{LiTCom}
			\end{figure*}

\re{Fig. \ref{LiTCom} depicts the LiTCom system model for image transmission in low-SNR conditions. 
At the transmitter, the source is coded into $\mathbf{s}$ according to \eqref{eq01}, which is a low-resolution image.  The pixel elements are partitioned based on sub-pixel importance (i.e., bit position significance \cite{xu2025dataimportanceJ}).  For the bit at the $k$-th position, the importance is modelled as its magnitude  $\gamma_k=2^{2(k-1)}$. This creates $K$ bit sub-sequences $\mathbf{s}_k$ of equal length. }

Each bit sub-sequence $\mathbf{s}_k$ is channel-encoded to produce a coded bit sub-stream $\mathbf{c}_k=\mathcal{F}_{\mathrm{ed\text{-}ch}}(\mathbf{s}_k)$. The considered error-distributing codes include: a) channel-uncoded transmission, and b)  weak FEC codes. These coded bit streams are subsequently modulated with the $M$-QAM modulation scheme:  
			\begin{equation}\label{eq03}
				\mathbf{x}_k = \mathcal{F}_{\mathrm{mod}}(\mathbf{c}_k),
			\end{equation}where $\mathcal{F}_{\mathrm{mod}}$ denotes the modulation function. The resulting modulated sub-stream $\mathbf{x}_k$ of length $L=S/\log_2(M)$ is normalized such that $\mathbb E[\mathbf{x}_k\mathbf{x}_k^\mathrm{H}]=\mathbf{I}_{L\times L}$ where $(\cdot)^\mathrm{H}$ denotes the Hermitian operation.

These modulated sub-streams are allocated with power according to an importance-aware strategy $\mathcal F_{\mathrm{pow}}$, aiming to extend uplink coverage (detailed in Sec. \ref{sec:V}). Assuming transmission over orthogonal sub-channels  \re{(e.g., orthogonal frequency-division multiplexing (OFDM) sub-carriers)}, the received signal of the $k$-th sub-stream is given by:
			\begin{equation}
				\mathbf{y}_{k} = h_k\sqrt{p_k}\mathbf{x}_k + \mathbf{n}_k,
			\end{equation}where $h_k$ is \re{sub-channel gain}, $p_k$ is the allocated transmit power, and $\mathbf{n}_k$ is additive Gaussian noise with each entry following $\mathcal{CN}(0,\sigma^2)$. The received SNR of the $k$-th sub-stream is given by:
			\begin{equation}
				\mathrm{snr}_k = \frac{p_k\vert h_k \vert^2}{\sigma^2}. 
			\end{equation}
			It is worth noting that the accurate, real-time estimation of $\mathrm{snr}_k$ is practically challenging \re{due to estimation inaccuracy, channel mismatch, and feedback delay.} 
			
			At the receiver, standard physical-layer operations are performed, including channel equalization $\mathcal H^{-1}_{\mathrm{equ}}$, demodulation $\mathcal{F}_{\mathrm{mod}}^{-1}$, and channel decoding $\mathcal{F}_{\mathrm{er\text{-}ch}}^{-1}$,  to obtain $\hat{\mathbf{x}}_k$,  $\hat{{\mathbf{c}}}_k$, and $\hat{\mathbf{s}}_k$, respectively. The received bit stream is given by:
			\begin{equation}
				\hat{\mathbf{s}}_k = \mathcal{F}_{\mathrm{er\text{-}ch}}^{-1}(\mathcal{F}_{\mathrm{mod}}^{-1}(\mathcal H^{-1}_{\mathrm{equ}}(\mathbf{y}_k)),
			\end{equation}which contains decoding errors, particularly under the low-SNR conditions. 
			
		For the channel-uncoded transimission, the BER of the $k$-th sub-stream is \cite{goldsmith2005wireless}:
			\begin{equation}\label{eq:BER_SNR_Uncoded}
				\mathrm{ber}_k^\mathrm{u} = \alpha_\mathrm{u} \mathcal{Q}\left(\beta_\mathrm{u}\sqrt{\mathrm{snr}_k}\right),
			\end{equation}where  $\alpha_\mathrm{u} =  \frac{4}{\log_2 M}(1-\frac{1}{\sqrt{M}})$, $\beta_\mathrm{u}=\sqrt{\frac{3}{M-1}}$, and  $\mathcal{Q}(x)=\frac{1}{\sqrt{2\pi}}\int_{x}^{\infty}\mathrm{exp}(\frac{-t^2}{2})dt$ is the Q-function. 
		
	 	 For weak FEC codes, the BER  of the $k$-th sub-stream  can be approximated as  \cite{xu2025dataimportanceJ}:
			\begin{equation}\label{eq:BER_SNR_Coded}
				\mathrm{ber}^\mathrm{c}_k \approx \alpha_\mathrm{c}\exp\left(\beta_\mathrm{c}\mathrm{snr}_k\right).
			\end{equation} $\alpha_\mathrm{c}> 0$ and $\beta_\mathrm{c}>0$ are parameters determined by the adopted channel coding and modulation scheme, which can be obtained via data fitting.

These bit errors across different sub-streams lead to the degradation in $\hat{\mathbf{s}}$. \re{By assuming at most one bit per pixel is incorrectly recovered,} the MSE between $\hat{\mathbf{s}}$ and $\mathbf{s}$ can be approximated by the importance-weighted MSE (IMSE) \cite{xu2025dataimportanceJ}:
\begin{align}\label{eq:MSE_BER}
	\mathrm{IMSE}(\hat{\mathbf{s}}, \mathbf{s}) &= \sum_{k}  \gamma_k \frac{\Vert \hat{\mathbf{s}}_{k}-\mathbf{s}_{k}\Vert^2 }{S} \approx \re{ \sum_{k}} \gamma_k\mathrm{ber}_k,
\end{align}where $\mathrm{ber}_k\in \{\mathrm{ber}_k^\mathrm{u}, \mathrm{ber}_k^\mathrm{c}\}$. 

 \section {\re{Perceived Uplink Coverage} \label{sec:IV}}
LiTCom aims to guarantee both perceptual and semantic fidelity in reconstructed content. Conventional distortion-based metrics, such as peak signal-to-noise ratio (PSNR), are inadequate for evaluation. This section introduces new evaluation metrics and defines \re{the perceived uplink coverage accordingly.}

\subsection{Evaluation Metrics\label{sec3.A}}
We propose to employ both NIQE and CLIP as complementary performance metrics to evaluate the perceptual and semantic quality, respectively.

	\subsubsection{NIQE-Based Perceptual Quality} 
	The NIQE metric, a non-reference image quality measure based on natural scene statistics \cite{mittal2012making}, provides an effective way to quantify image naturalness. Accordingly, we propose to evaluate the perceived quality using NIQE. Based on the discussion in Sec. \ref{sec:II}, it can be modelled as:
 	\begin{equation}
 		D_\mathrm{NIQE} (\mathrm{snr}, r) = \mathbb E \big[\mathrm{NIQE}(\hat{\mathbf{I}}_{\hat{\mathbf{s}}})\big],
 	\end{equation}where $\hat{\mathbf{I}}_{\hat{\mathbf{s}}}$ is reconstructed under the compression rate of $r$ and the channel condition of $\mathrm{snr}$, and  
 	\begin{equation}
 	\mathrm{NIQE}(\hat{\mathbf{I}}_{\hat{\mathbf{s}}})= \sqrt{(\boldsymbol{\mu}_{\hat{\mathbf{I}}_{\hat{\mathbf{s}}}} - \boldsymbol{\mu}_0)^T \left( \frac{\boldsymbol{\Sigma}_{\hat{\mathbf{I}}_{\hat{\mathbf{s}}}} + \boldsymbol{\Sigma}_0}{2} \right)^{-1} (\boldsymbol{\mu}_{\hat{\mathbf{I}}_{\hat{\mathbf{s}}}}- \boldsymbol{\mu}_0)}.
 	\end{equation} $\boldsymbol{\mu}_{\hat{\mathbf{I}}_{\hat{\mathbf{s}}}}, \boldsymbol{\Sigma}_{\hat{\mathbf{I}}_{\hat{\mathbf{s}}}}$ are the mean vectors and covariance matrices of the multivariate Gaussian (MVG) model of ${\hat{\mathbf{I}}}$, while $\boldsymbol{\mu}_0, \boldsymbol{\Sigma}_0$ correspond to pristine natural sources.
 	 Consistent with NIQE's scoring range, $D_\mathrm{NIQE} (r, \mathrm{snr}) \in [0,100]$, where lower values indicate higher perceptual quality.
  
 	\subsubsection{CLIP-Based Semantic Fidelity}However, good perceived quality does not necessarily mean that the semantic meaning of the source is successfully conveyed. For instance, in the presence of radio jamming,  the receiver may reconstruct a visually natural source yet a semantically irrelevant source.  To make the evaluation more comprehensive, we incorporate the CLIP model to measure the semantic similarity between the original source and its reconstruction \cite{xu2024semantic}. Similarly, the semantic fidelity metric is modelled as: 	\begin{align}
 		D_\mathrm{CLIP}&(\mathrm{snr}, r) = \mathbb{E} \big[ 1- \mathrm{{CLIP}}(\mathbf{I}, \hat{\mathbf{I}}_{\hat{\mathbf{s}}})\big],
 	\end{align}where:
 		\begin{equation}\label{eq07}
 		\mathrm{CLIP}(\mathbf{I}, \hat{\mathbf{I}}_{\hat{\mathbf{s}}}) = \frac{\phi(\mathbf{I})^\mathrm{T} \phi(\hat{\mathbf{I}}_{\hat{\mathbf{s}}})}{\|\phi(\mathbf{I})\|  \|\phi(\hat{\mathbf{I}}_{\hat{\mathbf{s}}})\|},
 	\end{equation}
 	where \( \phi(\cdot) \) denotes the CLIP model that transforms the inputs into an embedding vector \cite{radford2021learning}. It computes the cosine similarity between these two embedding vectors, producing $\mathrm{CLIP}(\mathbf{I}, \hat{\mathbf{I}}) \in [-1,1]$. Consequently,   $D_\mathrm{CLIP} \in [0,2]$, where lower values indicate higher semantic fidelity. 
  
   	$D_\mathrm{NIQE}$ and $D_\mathrm{CLIP}$ provide a complementary assessment of perceptual and semantic performance for LiTCom, both jointly related to $r$ and $\mathrm{snr}$. The overall QoE metric can be defined as:
   	\begin{equation}\label{eq:QoE}
   		\mathrm{QoE}(\mathrm{snr}, r) =   
   		[D_\mathrm{NIQE} (\mathrm{snr}, r), D_\mathrm{CLIP}(\mathrm{snr}, r)].
   	\end{equation}

	\subsection{\re{Perceived Uplink Coverage}} 
	
	\begin{figure}[tp]
		\centering
		\includegraphics[width=0.9\columnwidth]{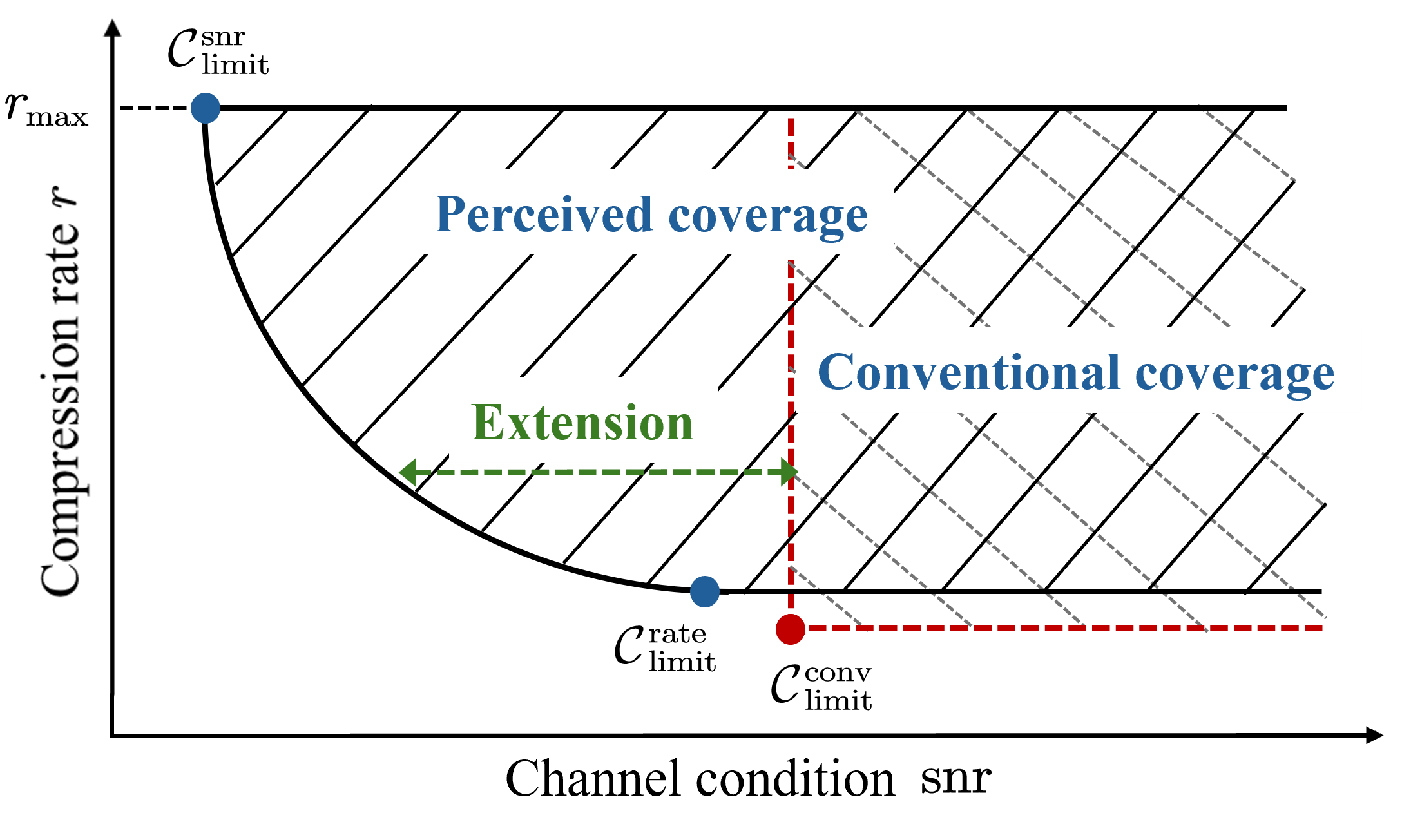}
		\caption{The comparison between the perceived coverage and conventional converage under a fixed uplink transmission rate assumption\re{, which aligns with the configured grant uplink mode specified in the 3GPP standards \cite{3gpp_ts_38_214}.} }
		\label{fig: coverage}
	\end{figure}
	 
	 Based on the proposed evaluation metrics, we introduce the concept of \emph{perceived uplink coverage}, which represents the region within which the reconstructed content satisfies perceptual and semantic qualities. \re{It is different from conventional coverage definition that is based on the bit-level reliability}. Formally, it can be quantified by the set of feasible $(r, \mathrm{snr})$ as follows:
	 \begin{equation}
	 	\mathcal C \triangleq \left\{ \left(\mathrm{snr}, r\right)\middle| \mathrm{QoE}(\mathrm{snr}, r) \preceq \mathrm{QoE}_\mathrm{th}\right\},
	 \end{equation}where  $\mathrm{QoE}_\mathrm{th}= [D_\mathrm{NIQE}^\mathrm{th}, D_\mathrm{CLIP}^\mathrm{th}]$, and $\preceq$ denotes the element-wise inequality. 

	Fig. \ref{fig: coverage} illustrates the concept of perceived coverage in contrast to the conventional one. The perceived coverage is upper bounded by $\bar{r}=1$, and lower bounded by the Rate-SNR function, defined as:  
			\begin{equation}
					\underline{r}(\mathrm{snr}) \triangleq \min r \quad\quad \mathrm{s.t.} \,\, \mathrm{QoE}(\mathrm{snr}, r) \preceq \mathrm{QoE}_{\mathrm{th}}.
			\end{equation}
 	Accordingly, two characteristic limit points can be identified: $\mathcal C_{\mathrm{limit}}^{\mathrm{snr}}\triangleq({\mathrm{snr}}^{\vdash},\underline{r}_{\mathrm{max}})$ and $\mathcal C_{\mathrm{limit}}^{\mathrm{rate}}\triangleq({\mathrm{snr}}^{\dashv},\underline{r}_{\mathrm{min}})$, where ${\mathrm{snr}}^{\vdash}$ and ${\mathrm{snr}}^{\dashv}$  are the minimum  SNRs required to satisfy the threshold $\mathrm{QoE}_\mathrm{th}$ at the compression rates of $\underline{r}_{\max}$ and $\underline{r}_{\min}$, respectively. 
	The limit point of the conventional coverage is represented by  $\mathcal C_{\mathrm{limit}}^{\mathrm{conv}}=({\mathrm{snr}}_{\mathrm{conv}}^{\vdash},\underline{r}_{\mathrm{conv}})$, where ${\mathrm{snr}}_{\mathrm{conv}}^{\vdash}$ and $\underline{r}_{\mathrm{conv}}$ are the minimum SNR and compression rate to guarantee the QoE requirement, respectively. 
	By harnessing the error robustness of LiTCom, the perceived coverage can be significantly extended. The gain can be characterized by:
			{\begin{equation}
					G = G_\mathrm{snr} + G_\mathrm{r},
			\end{equation}}where: 
			\begin{equation}
				G_\mathrm{snr}  = 10\log_{10}\left(\frac {\mathrm{snr}_\mathrm{conv}}{\mathrm{snr}}\right),\, G_\mathrm{r}  = 10\log_{10}\left(\frac{r_\mathrm{conv}}{r}\right).
			\end{equation} 
			
\section{Importance-Aware Power Allocation \label{sec:V}}
	The perceived uplink coverage of LiTCom can be further enhanced by  resource allocation. However, the explicit analytical form of the proposed QoE function in \eqref{eq:QoE} is non-trivial, making direct resource optimization intractable. To address this, we leverage the \re{assumption} that links the QoE and the MSE, which approximately characterizes the error robustness, and can be further approximated by IMSE in \eqref{eq:MSE_BER}. 
	The IMSE thus provides a tractable surrogate objective for the development of sub-optimal power allocation strategies. 
	
	Given the compression rate, the power allocation problem for QoE optimization can be approximately formulated as:		
			\begin{subequations}\label{eq:Prob_bit_level}
				\begin{align}
					 \min_{p_k} \quad &\sum_{k=1}^{K}\gamma_k\mathrm{ber}_{k}\\
					\mathrm{s.t.} \quad & \sum_{k=1}^{K} p_k \le P, \label{eq:cons_power}
				\end{align}
			\end{subequations}where  $P$ is the total power budget. $\mathrm{ber}_k \in \{\mathrm{ber}_k^{\mathrm u},   \mathrm{ber}_k^{\mathrm c}\}$ is the BER of the $k$-th sub-stream. 
			 
			Problem \eqref{eq:Prob_bit_level} is convex with respect to (w.r.t.)  the allocated power variable $p_k$. Therefore, the optimal solution can be obtained via the Lagrange multiplier technique  \cite{boyd2004convex}.  The Lagrangian function is: 	
			\begin{equation}
				\mathcal L(p_k, \lambda) \triangleq \sum_{k=1}^{K}\gamma_k\mathrm{ber}_{k} + \lambda\Big(\sum_{k=1}^{K} p_k -P\Big),
			\end{equation}where $\lambda\ge 0$ is the Lagrange multiplier. The optimal solution satisfies the following KKT conditions:
			\begin{subequations}
				\begin{align}
						&\frac{\partial \mathcal L(p_k, \lambda)}{\partial p_k}=\gamma_k\frac{\partial \mathrm{ber}_{k}}{\partial p_k} + \lambda=0,\label{eq:lagrange}\\
				&\lambda\Big(\sum_{k=1}^{K} p_k -P\Big) = 0. 
				\end{align}
			\end{subequations}In the following, we derive the corresponding solutions for both the channel-uncoded and channel-coded cases.
			
			\subsubsection{Channel-Uncoded Case} The gradient of the BER function $\mathrm{ber}_k^{\mathrm u}$ in   \eqref{eq:BER_SNR_Uncoded}  w.r.t. the allocated power $p_k$ is obtained as:  
			\begin{align}\label{eq:gradient_u}
				\frac{\partial \mathrm{ber}_{k}}{\partial p_k} = -\frac{1}{2\sqrt{\mathrm{snr}_k} }	\frac{\alpha_{\mathrm u}\beta_{\mathrm u}\vert h_k \vert^2}{\sqrt{2\pi}\sigma^2}\exp\Big(-\frac{\beta_{\mathrm u}^2 \mathrm{snr}_k}{2}\Big)<0,
			\end{align}which is strictly negative. This indicates, according to the KKT conditions, that the constraint in \eqref{eq:cons_power} holds with equality. 
			Substituting \eqref{eq:gradient_u} back into \eqref{eq:lagrange} yields: 
			\begin{align}\label{eq:gradient}
				   	\frac{\alpha_{\mathrm u}\beta_{\mathrm u}\vert h_k \vert^2}{\sqrt{2\pi}\sigma^2}\exp\Big(-\frac{\beta_{\mathrm u}^2 \mathrm{snr}_k}{2}\Big)=\frac{2\lambda}{\gamma_k}\sqrt{\mathrm{snr}_k},
			\end{align}which is a transcendental equation in terms of $p_k$. Since a closed-form expression of $p_k$ is unavailable for a given Lagrange multiplier $\lambda$, numerical methods are often used for root finding. In the outer loop, the optimal multiplier $\lambda^*$ is typically found through iterative search to satisfy the equality of \eqref{eq:cons_power}, \re{but incurring extremely high computational complexity due to both root-finding and multiplier updates.} 
			
			To reduce computational overhead, we propose a low-complexity  method by relaxing  \eqref{eq:gradient} into:
			\begin{align}\label{eq:gradient_u_appr}
				\frac{\alpha_{\mathrm u}\beta_{\mathrm u}\vert h_k \vert^2}{\sqrt{2\pi}\sigma^2}\exp\Big(-\frac{\beta_{\mathrm u}^2 \mathrm{snr}_k}{2}\Big)=\frac{2\lambda}{\gamma_k}\sqrt{\widetilde{\mathrm{snr}}_k},
			\end{align}where $\widetilde{\mathrm{snr}}_k\triangleq \frac{\tilde{p}_k\vert h_k \vert^2}{\sigma^2}$ and  $\tilde{p}_k \triangleq \frac{\gamma_kP}{\sum_k \gamma_k}$. Since allocated power is non-negative, the sub-optimal solution is obtained as:
			\begin{align}\label{eq:allop_bit_level_appr}
				p_k^* = \underbrace{\frac{2\sigma^2}{\beta_{\mathrm u}^2\vert h_k \vert^2}}_{W_k^{\mathrm u}}\Bigg(\underbrace{\ln \frac{\alpha_{\mathrm u}\beta_{\mathrm u}}{2\sqrt{2\pi}\lambda^*} }_{H^{\mathrm u*}_\mathrm{level}} - \underbrace{\ln\frac{\sigma^2\sqrt{\widetilde{\mathrm{snr}}_k}}{\gamma_k\vert h_k \vert^2}}_{H_k^{\mathrm u}} \Bigg)^+,
			\end{align}where $(\cdot)^+$ denotes the  $\max(0,\cdot)$ operation, and $\lambda^*$ is Lagrange multiplier solution. 
			
			This resembles the classical water-filling (WF) solution, where $W_{k}^{\mathrm u}$, $H_{k}^{\mathrm u}$, and $H_\mathrm{level}^{^{\mathrm u*}}$  can be interpreted as the base widths, base heights, and the water level associated with $\lambda^*$, respectively. The solution should satisfy the equality of \eqref{eq:cons_power}, leading to: 
			 			 \begin{equation}
			 	 \sum_{k=1}^{K} W_k^{\mathrm u} \left({H^{{\mathrm u*}}_\mathrm{level}} - {H_k^{\mathrm u}} \right)^+ - P = 0.
			 \end{equation} 
	$H_\mathrm{level}^{\mathrm u*}$ can be efficiently obtained via the bisection search technique.  Define the function $
			f({H}_\mathrm{level}^{\mathrm u}) = \sum_{k=1}^{K} W_k^{\mathrm u} \left({H^{{\mathrm u}}_\mathrm{level}} - {H_k^{\mathrm u}} \right)^+$,
	 which is monotonically increasing w.r.t. ${H}_\mathrm{level}^{\mathrm u}$. To perform the bisection search,   appropriate lower and upper bounds must first be established.  
	 
	 The lower bound, denoted by $\underline{H}_\mathrm{level}^{\mathrm u}$, is required to ensure $f(\underline{H}_\mathrm{level}^{\mathrm u})<P$, and can be set as $\underline{H}_\mathrm{level}^{\mathrm u} \triangleq \min_k(H_k^{\mathrm u})$. For the upper bound, denoted as $\overline{H}_\mathrm{level}^{\mathrm u}$, the condition $f( \overline{H}_\mathrm{level}^{\mathrm u})\ge P$ must hold. Given that $\overline{H}_\mathrm{level}^{\mathrm u}>\max_k({H_k^{\mathrm u}})$, this can be ensured by: 
	 		\begin{align}
			f( \overline{H}_\mathrm{level}^{\mathrm u}) & \ge  \min_k (W_k^{\mathrm u}) \sum_{k=1}^{K}  ({\overline{H}^{{\mathrm u}}_\mathrm{level}} - {H_k^{\mathrm u}})\nonumber\\
			& \ge  K \min_k (W_k^{\mathrm u}) ({\overline{H}^{{\mathrm u}}_\mathrm{level}} - \max_k({H_k^{\mathrm u}}))\ge P.
		\end{align}Thus, the upper bound can be set as $\overline{H}_\mathrm{level}^{\mathrm u}\triangleq \frac{P}{K\min_k (W_k^{\mathrm u})}+\max_k({H_k^{\mathrm u}})$. The procedure for obtaining ${H^{{\mathrm u*}}_\mathrm{level}}$  is summarized in \textbf{Algorithm}  \ref{alg:alg1}. The computational complexity is $\mathcal O({K\log_2\epsilon})$, where $\epsilon$ is the bisection searching accuracy.

		\begin{algorithm}[t] 
			\renewcommand{\algorithmicrequire}{\textbf{Input:}}
			\renewcommand{\algorithmicensure}{\textbf{Output:}}
			\caption{Importance-Aware Power Allocation for Channel-Uncoded Case.} 
			\label{alg:alg1}
			\begin{algorithmic}[1]		
				
				\STATE  Calculate the lower and upper bounds: 
				\[\underline {H}_{\mathrm{level}}^{\mathrm u}=\min_k(H_k^{\mathrm u}), \,\, \overline {H}_{\mathrm{level}}^{\mathrm u}=\frac{P}{K\min_k (W_k^{\mathrm u})}+\max_k({H_k^{\mathrm u}})\]
				\STATE \textbf{While} $\overline {H}_{\mathrm{level}}^{\mathrm u} - \underline {H}_{\mathrm{level}}^{\mathrm u}\ge \epsilon$
				\STATE  \quad Set new $ {H}_{\mathrm{level}}^{\mathrm u} \leftarrow (\overline {H}_{\mathrm{level}}^{\mathrm u} + \underline {H}_{\mathrm{level}}^{\mathrm u})/2$
				\STATE  \quad Compute $p_k$ based on (\ref{eq:allop_bit_level_appr})	
				\STATE  \quad \textbf{If} $\sum_{k=1}^{K}p_k \le P$
				\STATE  \quad \quad Update $\underline {H}_{\mathrm{level}}^{\mathrm u}\leftarrow {H}_{\mathrm{level}}^{\mathrm u}$
				\STATE  \quad \textbf{Else}
				\STATE  \quad \quad  Update $\overline {H}_{\mathrm{level}}^{\mathrm u}\leftarrow {H}_{\mathrm{level}}^{\mathrm u}$
				
				\STATE  \quad \textbf{End}	
				\STATE  \textbf{End}
				\STATE   Compute $p_k^*$ based on (\ref{eq:allop_bit_level_appr})	
			\end{algorithmic}  		
		\end{algorithm}
		
		\subsubsection{Channel-Coded Case}The power allocation strategy for the channel-coded case can be developed in a similar manner. Following our previous work \cite{xu2025dataimportanceJ}, the optimal power allocation is given by:
			\begin{equation}\label{eq:allop_coded}
				p_k^* = \underbrace{-\frac{\sigma^2}{\beta_{\mathrm c} \vert h_k \vert^2}}_{W_k^{\mathrm c}}\Big(\underbrace{\ln \frac{-\alpha_{\mathrm c}\beta_{\mathrm c}}{\lambda^*} }_{H^{\mathrm c*}_\mathrm{level}} - \underbrace{\ln\frac{\sigma^2}{\gamma_k\vert h_k \vert^2}}_{H_k^{\mathrm c}} \Big)^+.
			\end{equation} The optimal water level $H_\mathrm{level}^{\mathrm c*}$ corresponds to the optimal Lagrange multiplier $\lambda^*$ and satisfies the quality of \eqref{eq:cons_power}:
			\begin{equation}
				\sum_{k=1}^{K} {W_k^{\mathrm c}}\left({H^{\mathrm c*}_\mathrm{level}} - {H_k^{\mathrm c}} \right)^+ = P,
			\end{equation}which can be solved via the bisection search technique. Similarly, the lower bound  can be set to
			$\underline{H}_\mathrm{level}^{\mathrm c} \triangleq \min_k(H_k^{\mathrm c})$, ensuring $\sum_{k=1}^{K} {W_k^{\mathrm c}}\left({H^{\mathrm c*}_\mathrm{level}} - {H_k^{\mathrm c}} \right)^+<P$. The upper bound can be chosen as $\overline{H}_\mathrm{level}^{\mathrm c}\triangleq \frac{P}{K\min_k (W_k^{\mathrm c})}+\max_k({H_k^{\mathrm c}})$ to guarantee $\sum_{k=1}^{K} {W_k^{\mathrm c}}\left({H^{\mathrm c*}_\mathrm{level}} - {H_k^{\mathrm c}} \right)^+>P$.  The procedure for solving the optimal power allocation $p_k^*$  for the channel-coded case follows a similar pattern to that of the channel-uncoded case with a computational complexity of $\mathcal O(K\log_2 \epsilon)$, which is summarized in \textbf{Algorithm}  \ref{alg:alg2}.
			
			\begin{algorithm}[t] 
				\renewcommand{\algorithmicrequire}{\textbf{Input:}}
				\renewcommand{\algorithmicensure}{\textbf{Output:}}
				\caption{Importance-Aware Waterfilling for Channel-coded Case.} 
				\label{alg:alg2}
				\begin{algorithmic}[1]		
					\STATE  Calculate the lower and upper bounds: \[\underline {H}_{\mathrm{level}}^{\mathrm c}=\min_k(H_k^{\mathrm c}), \overline {H}_{\mathrm{level}}^{\mathrm c}= \frac{P}{K\min_k (W_k^{\mathrm c}) }+\max_k({H_k^{\mathrm c}})\] 
					
					\STATE \textbf{While} $\overline {H}_{\mathrm{level}}^{\mathrm c} - \underline {H}_{\mathrm{level}}^{\mathrm c}\ge \epsilon$
					\STATE  \quad Set new $ {H}_{\mathrm{level}}^{\mathrm c} \leftarrow (\overline {H}_{\mathrm{level}}^{\mathrm c} + \underline {H}_{\mathrm{level}}^{\mathrm c})/2$
					\STATE  \quad Compute $p_k$ based on (\ref{eq:allop_coded})	
					\STATE  \quad \textbf{If} $\sum_{k=1}^{K}p_k \le P$
					\STATE  \quad \quad Update $\underline {H}_{\mathrm{level}}^{\mathrm c}\leftarrow {H}_{\mathrm{level}}^{\mathrm c}$
					\STATE  \quad \textbf{Else}
					\STATE  \quad \quad  Update $\overline {H}_{\mathrm{level}}^{\mathrm c}\leftarrow {H}_{\mathrm{level}}^{\mathrm c}$
					
					\STATE  \quad \textbf{End}	
					\STATE  \textbf{End}
					\STATE   Compute $p_k^*$ based on (\ref{eq:allop_coded})	
				\end{algorithmic}  			
			\end{algorithm}
		
			\section{Simulation Results \label{sec:VI}}
 
				This section presents simulation results to validate the effectiveness of LiTCom for uplink transmission under low-SNR conditions. The visual source considered is an RGB image with a resolution of $H=640$, $W=512$, and an $8$-bit depth. 
				\subsection{Parameter Setup}
				At the transmitter, the semantic-preserving source encoder applies a low-pass mean filter to produce a compressed representation of the input image, which is partitioned into $K=8$ bit sequences based on the sub-pixel-level importance. Under low-SNR conditions, the error-distributing codes are considered: channel-uncoded transmission, a $(7,4)$ Hamming code, and a rate $1/2$ convolutional code with constraint length of $7$ and generator polynomials of $[171, 133]_8$. The modulation order is set to $M=4$.  At the receiver, the offline-trained SUPIR model is employed as the generative decoder \cite{yu2024scaling}. \re{The LiTCom transmitter pipeline produces received representations whose distortions fall within this pretraining regime. The LPF-based source compression produces low-resolution distortion, while channel-uncoded transmission at low SNR produces approximately i.i.d. residual errors that resemble the low-quality degradation of the training dataset. LiTCom is therefore robust, conditioning on the residual errors remaining within this training degradation regime of the GenAI model.}
				
				The QoE requirement is defined as $\mathrm{QoE}_\mathrm{th}= [D_\mathrm{NIQE}^\mathrm{th}, D_\mathrm{CLIP}^\mathrm{th}]=[5, 0.1]$.   
				\re{Under orthogonal sub-channels, fading can be represented by parallel AWGN sub-channels with different effective post-equalization SNRs. The AWGN performance curves therefore characterize the conditional behavior of each sub-channel given its instantaneous SNR, and the performance under any specific fading distribution can be obtained by averaging these curves over the corresponding SNR distribution \cite{simon2004digital}. In rapidly changing channels, CSI errors introduce residual interference into the equalized signal. With MMSE estimation, this effect can be folded into a signal-to-interference-plus-noise ratio (SINR) and approximately interpreted as an effective-SNR reduction.  Accordingly, the AWGN model is used to assess the proposed LiTCom design.}
			 
				\re{To benchmark the proposed LiTCom framework, we consider both the 5G NR-like and Deep-JSCC baselines. Since JPEG is one of the most widely used image compression standards \cite{wallace1992jpeg}, and LDPC is adopted for the data channel in 5G NR \cite{3gpp_ts_38212}, the 5G NR-like system baseline adopts JPEG for source coding and  LDPC for channel coding. The LDPC adopts BG1 with information block length of 8448 and the coding rate of 1/3, while channel-uncoded case is also considered as a conventional approach.}  The representative Deep-JSCC \cite{bourtsoulatze2019deep} is also compared, which is trained under specific channel conditions and compression rates. Equal power (EP) allocation strategy is applied in all baseline schemes to highlight the exploitation of data importance. Conventional WF ignores such importance and only adapts fading channels, which is reduced into EP allocation under AWGN channels.  The following approaches are evaluated:
				\begin{itemize}
					\item  GenAI-Uncoded-WF/GenAI-Uncoded-EP: Proposed LiTCom framework with uncoded transmission using importance-aware WF  or  EP allocation.
	 
					\item GenAI-Hamming-WF/GenAI-Hamming-EP: Proposed LiTCom framework with Hamming code   using importance-aware WF or EP allocation.
			 
					\item GenAI-Conv.-WF/GenAI-Conv.-EP: Proposed LiTCom framework with  convolutional code   using importance-aware WF or  EP allocation.
			 
					\item  JPEG-Uncoded-EP/JPEG-LDPC-EP: Conventional communication baselines without channel coding or with LDPC.
					
					\item Deep-JSCC \cite{bourtsoulatze2019deep}: Classical semantic paradigm baseline based on autoencoder network.		
				\end{itemize}
							 
\subsection{Effectiveness of LiTCom Framework} 
 \begin{figure}[tp]
					\centering
					\subfigure[]{\includegraphics[width=0.225\textwidth]{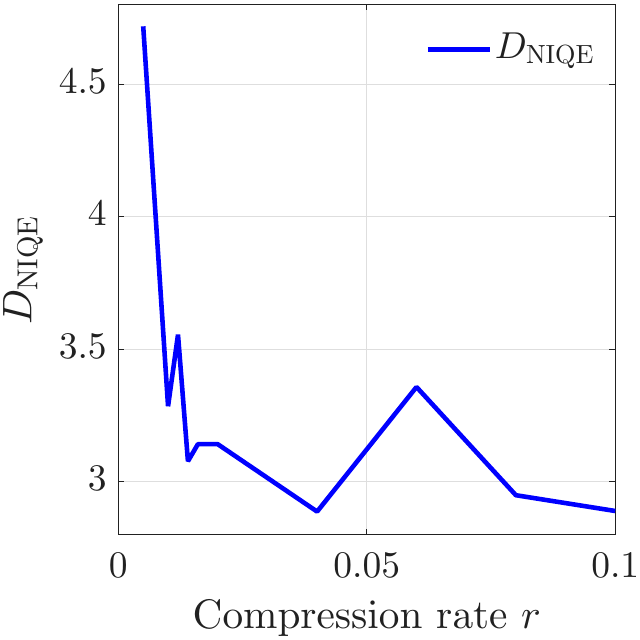}}		 	
					\hspace{5pt}
					\subfigure[]{\includegraphics[width=0.23\textwidth]{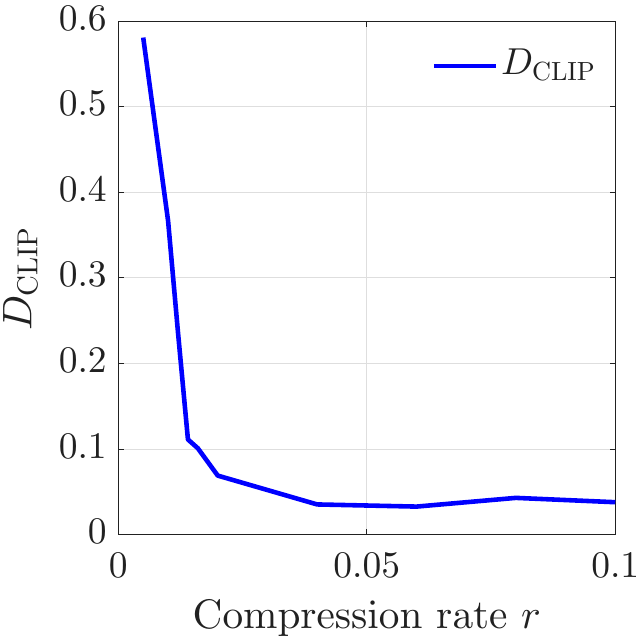}}
					\caption{Representing sufficiency of LiTCom under perfect transmission.}
					\label{fig:sufficiency}
				\end{figure}		
				We first consider error-free transmission, i.e., $\hat{\mathbf{s}}=\mathbf{s}$, to evaluate the proposed source coding scheme.  Fig. \ref{fig:sourceCoding} shows the original sources, their compressed representations, and the reconstructed outputs  evaluated by PSNR, SSIM, LPIPS, and CLIP similarity. They improve as the compression rate $r$ increases, indicating that the representation contains more sufficient semantic information for generative decoding. Then, we evaluate the proposed error-distributing channel codes at a compression rate of $r=9\%$.  Fig. \ref{fig:channelCoding}  depicts the reconstructed images at $E_b/N_o=-2$ dB, where the Deep-JSCC model is trained at $E_b/N_o=0$ \re{on CIFAR10 dataset}. The conventional 5G NR-like system fails because JPEG introduces strong inter-symbol dependency and LDPC decoding degrades at low SNR. Deep-JSCC achieves the highest PSNR and SSIM scores, as it is  trained to minimise MSE,  but its visual quality is less satisfactory under the proposed QoE metric. In contrast, GenAI-Uncoded-WF achieves the best NIQE and CLIP similarity, producing more natural and semantically consistent reconstructions. This confirms that the proposed NIQE-CLIP QoE metric better reflects human-perceived reconstruction quality.

				
				
				Fig.  \ref{fig:sufficiency}  examines the representing sufficiency of the  LPF-based semantic-preserving source encoder. \re{The QoE improves with increasing compression rate, and the representation becomes sufficient when $r>2\%$.} 
				Fig.~5 evaluates error robustness under channel-uncoded transmission with EP allocation. {\color{black}The errors between $s$ and $\hat{s}$ are quantified using the MSE normalized by the average power per pixel per channel.} The QoE degrades as the normalized MSE increases, but remains satisfactory when the MSE is below $0.025$, \re{which supports Assumption~1. } The channel-uncoded configuration achieves the best QoE performance, followed by the Hamming-coded and the convolutional-coded cases. As is shown in Fig.~\ref{fig:distributingcode}, this is because channel-uncoded transmission produces more independently distributed errors, whereas channel coding tends to introduce more structured and correlated error patterns, \re{supporting  \textbf{Assumption~2}.} Under the same error-distributing code, the importance-aware WF strategy achieves lower MSE and better QoE performance compared to the EP strategy.   
	 
	 \begin{figure}[tp]
	\centering
	\subfigure[]{\includegraphics[width=0.24\textwidth]{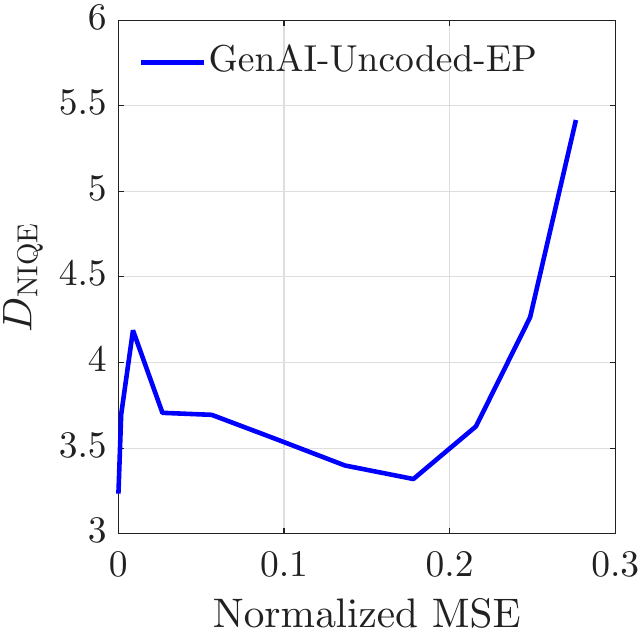}}
	\subfigure[]{\includegraphics[width=0.24\textwidth]{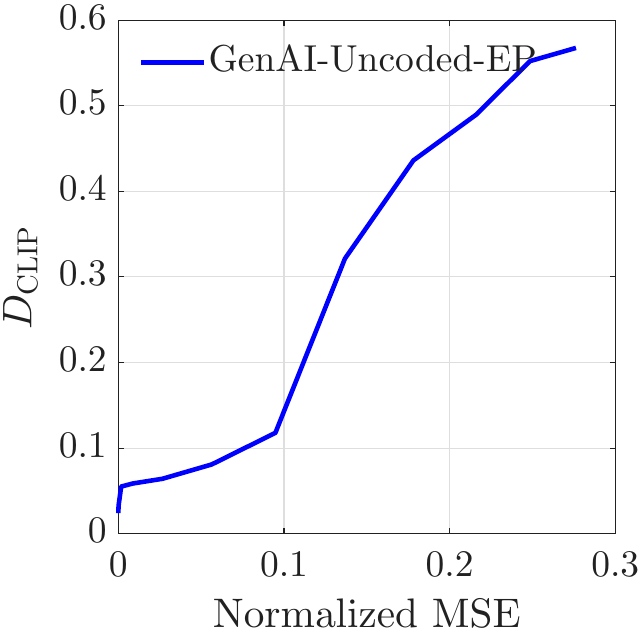}}	
	\caption{Error robustness of LiTCom given the LPF-based semantic-preserving encoder with a compression rate of $9\%$.}
	
	\label{fig:robustness} 
\end{figure}

\begin{figure}[tp]
	\centering
	\subfigure[]
	{\includegraphics[width=0.44\textwidth]{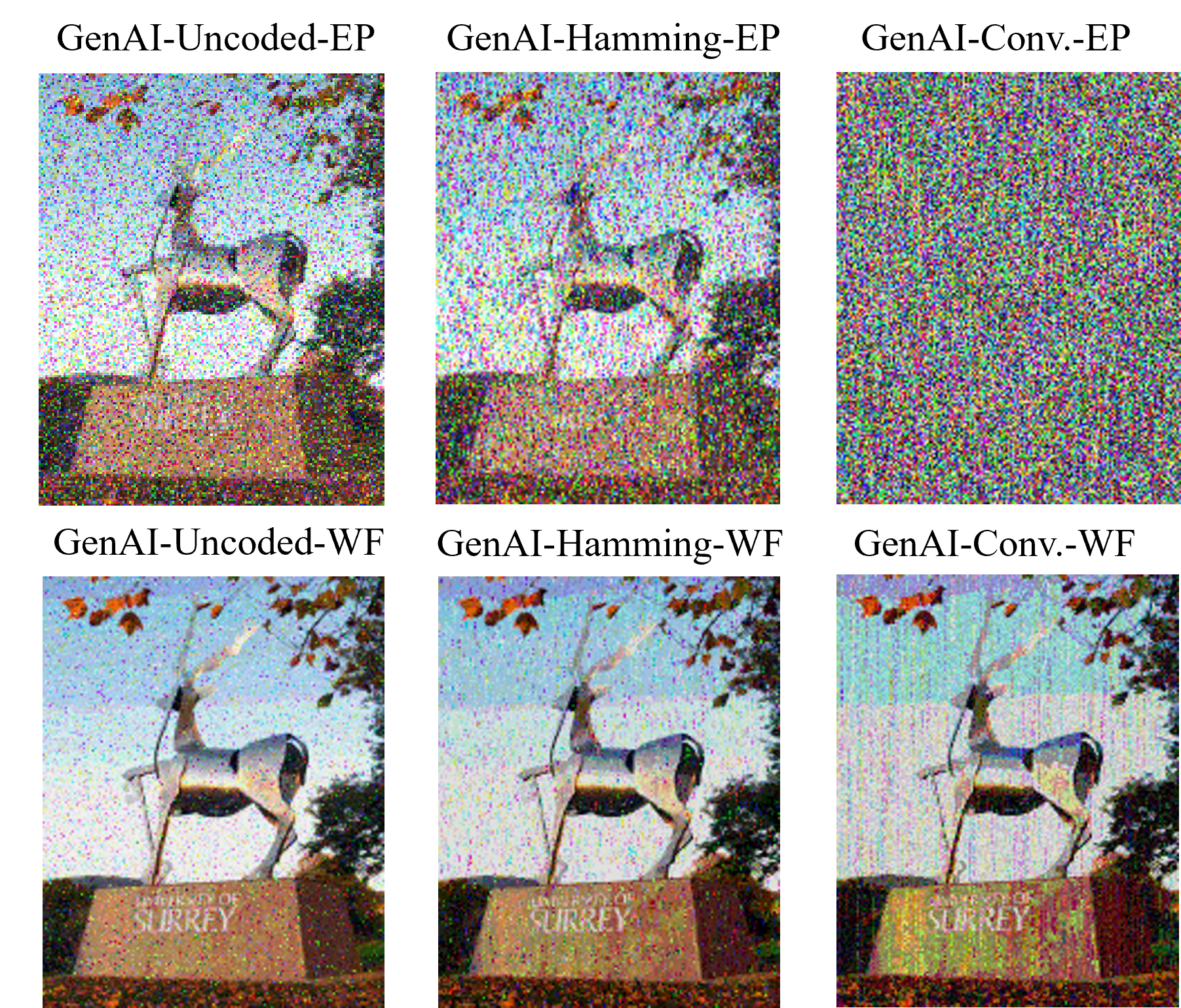}}	
	\caption{Error distribution across the source representations at a compression rate of $9\%$ and the channel condition of $E_b/N_o=-2$ dB.}
	
	\label{fig:distributingcode} 
\end{figure}

\subsection{Performance Comparison}
\begin{figure*}
	 \centering
	 \includegraphics[width=1\textwidth]{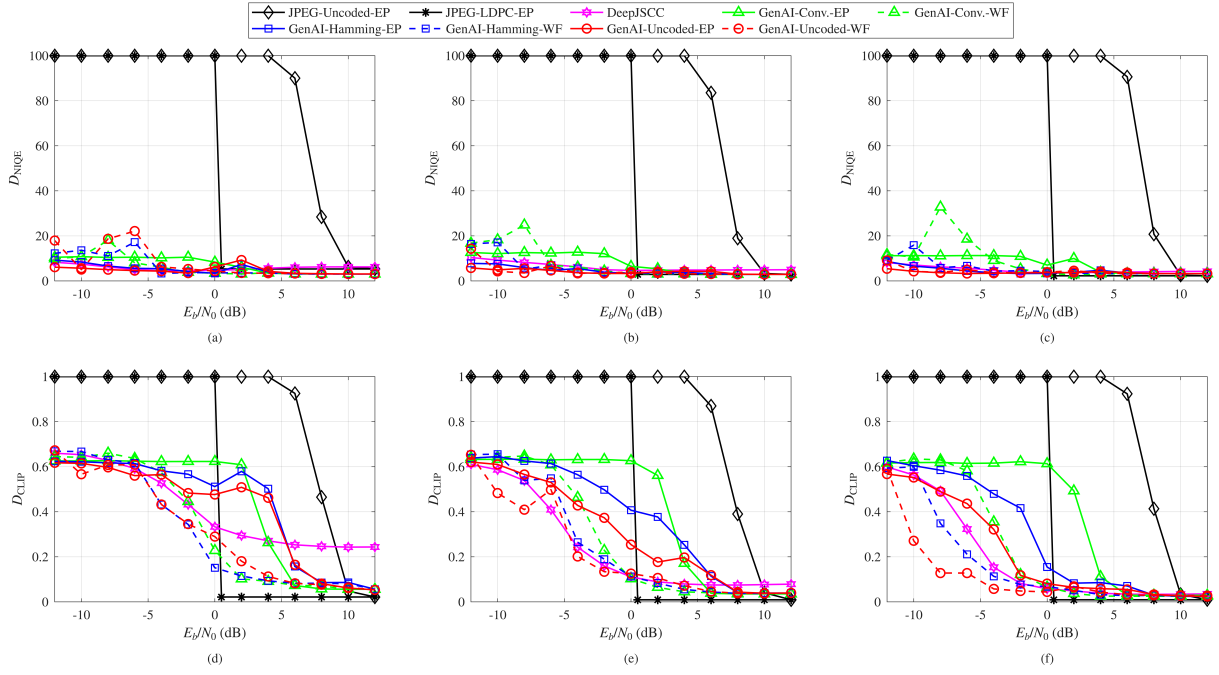}
	 \caption{The QoE Performance comparison: (a) $D_\mathrm{NIQE}$ at a compression rate of $r=2\%$, (b) $D_\mathrm{NIQE}$ at a compression rate of $r=4\%$, (c) $D_\mathrm{NIQE}$ at a compression rate of $r = 9\%$, (d) $D_\mathrm{CLIP}$ at a compression rate of $r=2\%$, (e) $D_\mathrm{CLIP}$ at a compression rate of $r=4\%$, (c) $D_\mathrm{CLIP}$ at a compression rate of $r = 9\%$.}
	 \label{fig:QoE_Comparison}
\end{figure*}

	\begin{figure}[tp]
		\centering
		\includegraphics[width=0.4\textwidth]{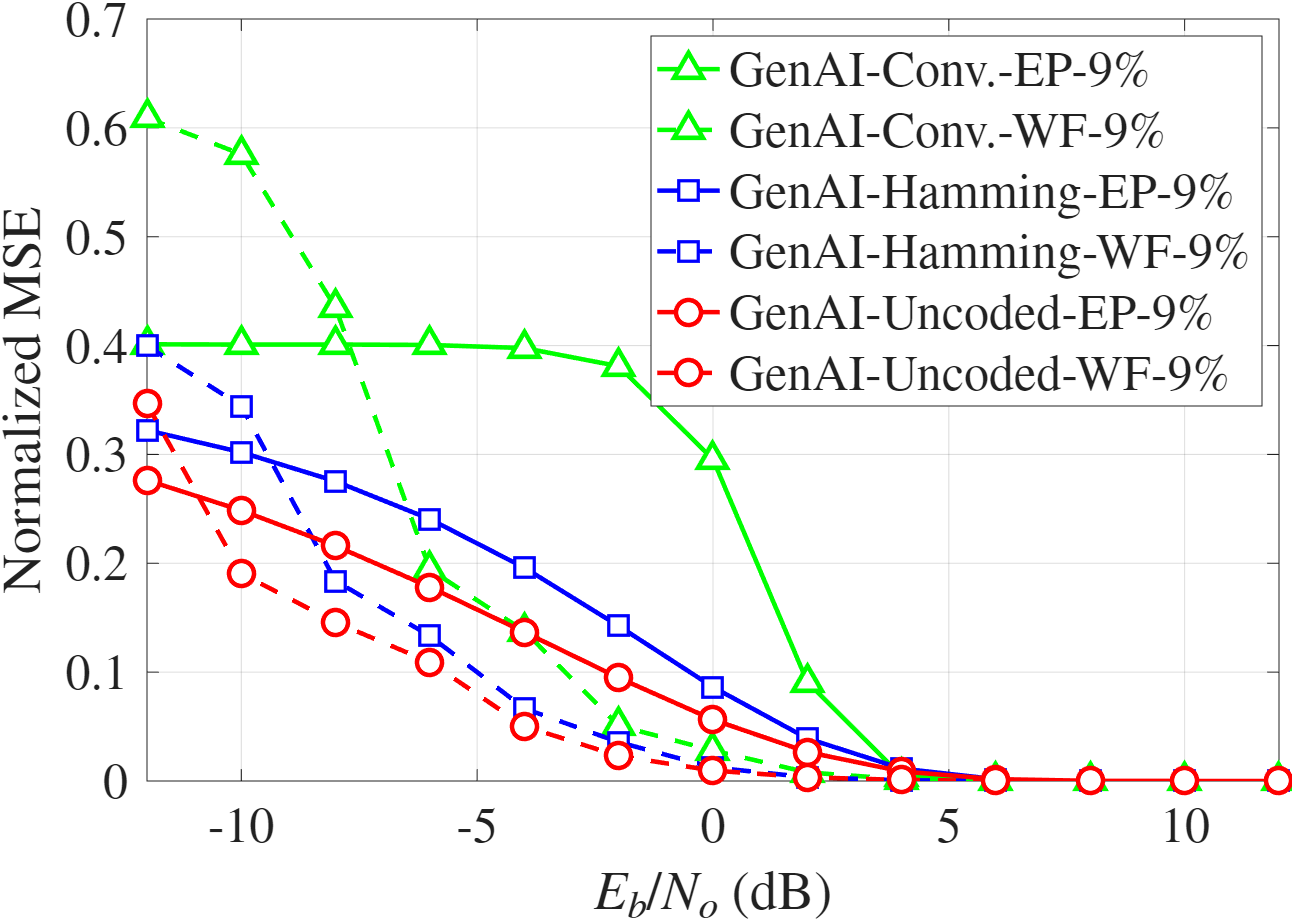}
		\caption{Performance comparison in terms of normalized MSE under the compression rate of  $9\%$.}
		\label{fig:MSE_SNR}
	\end{figure} 
\begin{figure}[tp]
	\centering
 	\includegraphics[width=0.8\columnwidth]{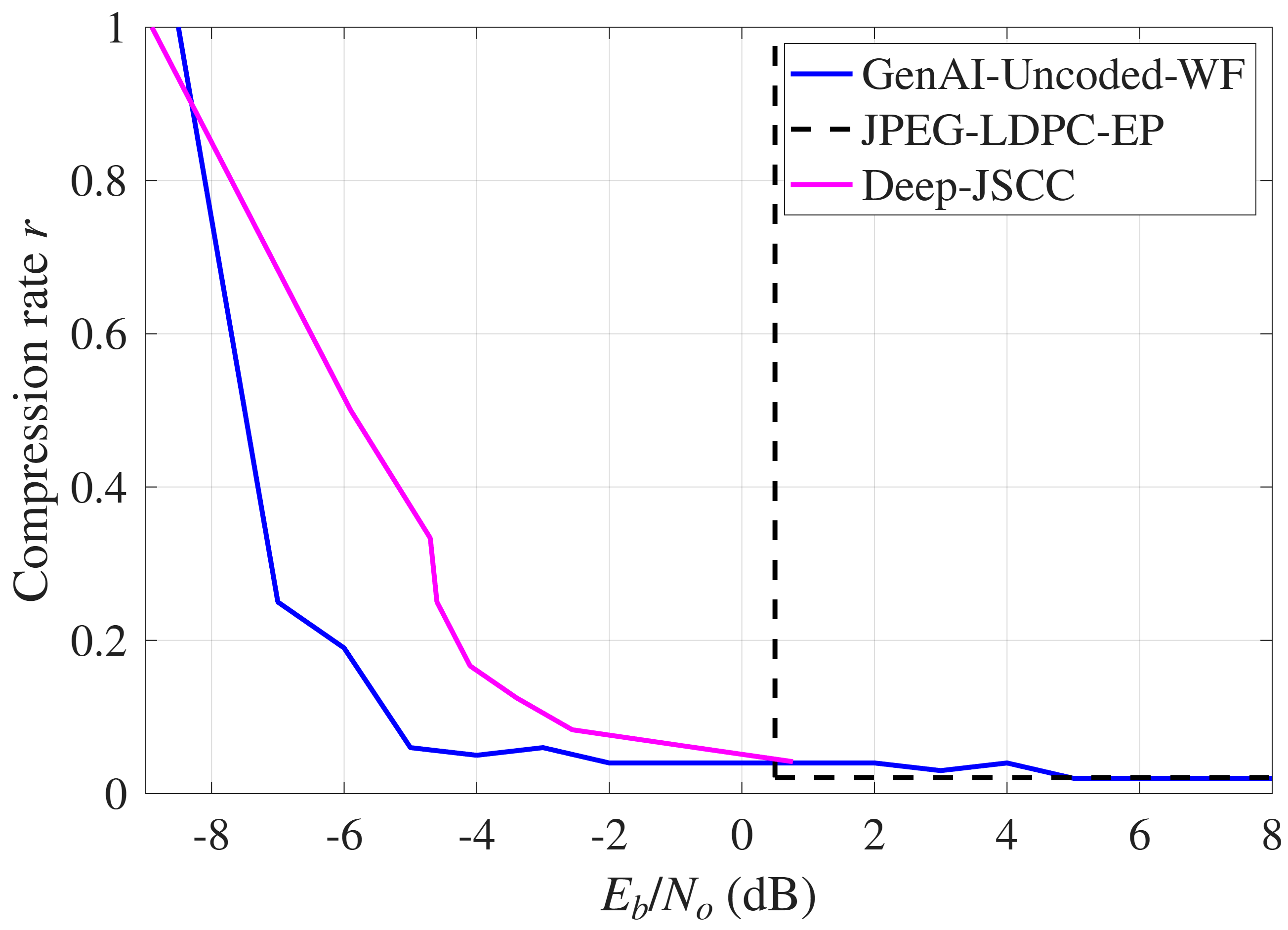}
	\caption{{Perceived coverage of LiTCom in comparison with traditional wireless system and classical Deep-JSCC model given a fixed uplink transmission rates}}
	\label{fig:perceivedCoverage}
\end{figure}

	   Fig. \ref{fig:QoE_Comparison}  compares QoE performance of different schemes. Under low SNR conditions, 5G NR-like baseline suffers severe degradation, with QoE performance of $D_\mathrm{NIQE}=100$ and $D_\mathrm{CLIP}=1$.  The JPEG-Uncoded-EP and JPEG-LDPC-EP baselines  fail to meet the QoE requirement when  $E_b/N_o\le 4$ dB and $E_b/N_o\le 0$ dB, respectively. At  $r=9\%$, the required SNR of the GenAI-Uncoded-WF scheme is reduced by $14$ dB and $4.5$ dB, respectively.   $D_\mathrm{NIQE}$  and $D_\mathrm{CLIP}$  generally improve as  SNR increases under both the LiTCom framework and the Deep-JSCC baseline.   LiTCom shows minor fluctuations due to the distribution mismatches between the test images and training datasets.  Deep-JSCC meets the QoE requirements when  $E_b/N_o\le 2$ dB and $E_b/N_o\le -2$ dB at the compression rates of $4\%$ and $9\%$, respectively, but fails at $r=2\%$.

	   At low SNRs (i.e., $E_b/N_o\le 0$), channel-uncoded transmission achieves the best QoE performance, followed by Hamming and  convolutional coding. This is because the residual decoding errors become progressively more structured and correlated, causing the decoded representations to deviate from the GenAI training distribution. \re{The same mechanism explains the potential impact of burst errors caused by time-selective fading or decoding failures of strong FEC codes.} 
	  At higher SNRs, the convolutional-coded scheme eventually becomes preferable due to its stronger error-correction capability. These results suggest that LiTCom can be further improved by adapting the GenAI decoder to channel-error-contaminated representations and by selecting error-distributing codes according to the SNR regime.

	The required SNR decreases as the compression rate increases in LiTCom, whereas it remains nearly unchanged for conventional baselines. Such observation indicates that increased source redundancy enhances robustness to channel impairments and highlights the coupling between source and channel coding. 
	  Fig. \ref{fig:MSE_SNR} further demonstrates that the proposed importance-aware WF approach consistently outperforms EP by reducing the MSE. 
	  \re{Note that rapid channel variations may cause CSI errors, whose impact can be approximately captured as an effective-SNR reduction. 
	  Thus, the low-SNR results in Fig.~7 and Fig.~8 provide an indicative view of the performance of the two power allocation strategies under rapidly changing channels. For WF schemes, CSI errors may cause power misallocation and further degrade performance compared with the EP strategies. However, the proposed strategy can partly mitigate this effect by accounting for semantic importance, rather than relying solely on channel gains as in classical WF.}  

	To demonstrate the perceived coverage extension, we compare the proposed GenAI-Uncoded-WF  scheme  with the conventional JPEG-LDPC-EP and Deep-JSCC baselines under a fixed transmission rate. As depicted in Fig.~\ref{fig:perceivedCoverage}, 	the proposed LiTCom framework significantly extends the uplink perceived coverage. Specifically, the JPEG-LDPC-EP baseline meets the QoE requirement only when \( E_b/N_o \geq 0.5 \) dB and \( r \geq 2.1\% \), corresponding to a limit point of \( \mathcal{C}_{\mathrm{limit}}^{\mathrm{conv}} = (0.5\,\mathrm{dB},\,2.1\%) \). The Deep-JSCC is characterized by the limit points  \( \mathcal{C}_{\mathrm{limit}}^{\mathrm{snr}} = (-8.9\,\mathrm{dB},\,100\%) \)  and \( \mathcal{C}_{\mathrm{limit}}^{\mathrm{rate}} = (2\,\mathrm{dB},\,4.2\%) \). In contrast, the limit points of LiTCom are \( \mathcal{C}_{\mathrm{limit}}^{\mathrm{snr}} = (-8.5\,\mathrm{dB},\,100\%) \) and \( \mathcal{C}_{\mathrm{limit}}^{\mathrm{rate}} = (5\,\mathrm{dB},\,2\%) \) for the WF strategy, and \( \mathcal{C}_{\mathrm{limit}}^{\mathrm{snr}} = (-9\,\mathrm{dB},\,100\%) \) and \( \mathcal{C}_{\mathrm{limit}}^{\mathrm{rate}} = (8\,\mathrm{dB},\,2\%) \) for the EP approach. 
	Within the compression rate range of $ 33\%\sim 50\%$, the uplink coverage gain remains around $7.5\sim 8$ dB over JPEG-LDPC-EP and $2\sim 2.5$ dB over Deep-JSCC. This highlights LiTCom's strong robustness under low-SNR channel conditions, with more visualized examples provided in Fig.~\ref{fig:robustChannel} (in the appendix).

	\subsection{\re{Complexity, Latency, and Practical Deployment}}	
	
	\begin{table*}[t]
		\centering
		\caption{\normalsize \re{Computational complexity comparison.}}
		\label{tab:1}
		
		\begin{tabular}{|l|c|c|c|c|c|}
			\hline
			\textbf{Scheme} 
			& \textbf{Source Encoding } 
			& \textbf{Channel Encoding} 
			& \textbf{Channel Decoding} 
			& \textbf{Source Decoding} 
			& \textbf{Power Allocation} \\
			\hline
		 
			GenAI-Uncoded-WF/EP 
			& \re{$88$K Mults, $0.9$M Adds} 
			& 0 
			& 0 
			& $3.2$B Params 
			& $\mathcal O(K\log_2 \epsilon)$/ $\mathcal O(1)$ \\
			\hline
			GenAI-Hamming-WF/EP 
			& \re{$88$K Mults, $0.9$M Adds} 
			& $\mathcal O(N_\mathrm{Ham})$ 
			& $\mathcal O(N_\mathrm{Ham}\log_2 N_\mathrm{Ham})$ 
			& $3.2$B Params 
			& $\mathcal O(K\log_2 \epsilon)$/ $\mathcal O(1)$ \\
			\hline
			GenAI-Conv.-WF/EP 
			& \re{$88$K Mults, $0.9$M Adds} 
			& $\mathcal O(L_\mathrm{Cons}S)$ 
			& $\mathcal O(2^{L_\mathrm{Cons}-1}S)$ 
			& $3.2$B Params 
			& $\mathcal O(K\log_2 \epsilon)$/ $\mathcal O(1)$ \\
			\hline
			JPEG-LDPC-WF/EP 
			& \re{$2.7$M Mults, $7.1$M Adds} 
			& $\mathcal O(N_\mathrm{LDPC})$ 
			& $\mathcal O(I_\mathrm{iter}N_\mathrm{LDPC})$ 
			& $\mathcal O(I\log I)$ 
			& $\mathcal O(K\log_2 \epsilon)$/ $\mathcal O(1)$ \\
			\hline
			Deep-JSCC 
			& \multicolumn{2}{c|}{\makecell{$71.725$K Params}} 
			&  \multicolumn{2}{c|}{\makecell{$71.725$K Params}} 
			& N/A \\
			
			\hline
		\end{tabular}
	\end{table*}

\subsubsection{\re{Computational Complexity}}	
		Table \ref {tab:1} compares the computational complexity of LiTCom, the 5G NR-like baseline, and Deep-JSCC. At the transmitter, the proposed LPF-based source encoder requires \re{$88$K multiplications and $0.9$M additions, whereas  JPEG needs approximately $2.7$M multiplications and $7.1$M additions, mainly due to the DCT operation}. For channel coding,   channel-uncoded transmission incurs no computation, while Hamming encoding requires linear complexity of $\mathcal O(N_\mathrm{Ham})$ and convolutional coding of $\mathcal O(L_\mathrm{Cons}S)$, where  $N_\mathrm{Ham}$, $L_\mathrm{Cons}$, and $S$ are the codeword length, the constraint length and frame length, respectively.  In contrast, LDPC encoding introduces the complexity of $\mathcal O(N_\mathrm{LDPC})$, where $L_\mathrm{LDPC}$ denotes the codeword length. Deep-JSCC requires $3.1 G$ FLOPs at the NN encoder with $71.7 K$ parameters.  The computational complexity for power allocation strategies is the same for both LiTCom and the conventional baseline. 
		
	 		At the receiver, the deployed SUPIR consists of a Stable Diffusion XL backbone with approximately $2.6$ billion parameters and an adaptor module with $0.6$ billion parameters.  JPEG decoding has complexity of $\mathcal O(I\log I)$, while  Hamming and convolutional decoding require $\mathcal O(N_\mathrm{Ham}\log_2 N_\mathrm{Ham})$ per codeword and $\mathcal O(2^{L_\mathrm{Cons}-1}S)$ for Viterbi decoding per frame, respectively. LDPC decoding requires $\mathcal O(I_\mathrm{iter}N_\mathrm{LDPC})$, where $I_\mathrm{iter}$ is the number of belief propagation iterations. The Deep-JSCC decoder has comparable complexity to its encoder due to the symmetric nature of the autoencoder. 
		Overall, LiTCom reduces transmitter-side computation by more than 95\% compared with conventional baselines, making it suitable for resource-constrained 6G uplink devices.

\subsubsection{\re{Latency}}	 
	 \re{End-to-end latency and energy consumption are not used as the primary evaluation metrics in this work, since they depend strongly on hardware platforms, software implementation, and coding practice.  As an indicative measurement at $r=9\%$,  JPEG compression and reconstruction take $0.0468$ s and $0.0498$ s, respectively, while LDPC encoding and decoding take $0.0040$ s and $0.2430$ s in MATLAB R2025a on an Intel i5-14600K CPU with 128 GB RAM.  Deep-JSCC encoder and decoder inference each take $0.0016$ s on an NVIDIA GeForce RTX 5090 GPU.  For LiTCom, the LPF-based source coding takes $0.1072$ s on the same CPU platform. It is longer than 5G NR-like and Deep-JSCC baselines because JPEG compression benefits from the highly optimised implementation and Deep-JSCC encoder benefits from the GPU accelerator. The receiver-side generative inference takes approximately $1.2080$ s per diffusion step on the RTX 5090 GPU, corresponding to $30.2000$ s for 25 diffusion steps.  This inference latency  can be potentially reduced through adaptive-step inference according to received representation quality, model distillation, batching, and parallel execution on shared AI accelerators. These values are therefore reported only as implementation-level references, rather than intrinsic algorithmic limits. }
	 
	 \re{For wireless transmission, we further provide a normalized slot-level resource-occupancy estimate under NR numerology $\mu=1$, where the slot duration is $T_{\rm slot}=0.5$ ms. Assuming $70\%$ data resource element availability per resource block (RB), the estimated one-RB transmission times are $4499.0$ ms for the 5G NR-like baseline and $1500.0$ ms for uncoded LiTCom. Deep-JSCC requires 375.0 ms under the idealized assumption that each continuous-valued complex channel symbol occupies one OFDM resource element. These values compare minimum wireless resource occupancy under a common OFDM frame structure, rather than software processing time or full system-level scheduling latency. }  

\subsubsection{\re{Practical Deployment}}		 
	 \re{This receiver-side computational demand has direct implications for network deployment. LiTCom should not be interpreted as requiring every  BS sector to independently execute a large diffusion model for all users in real time. A more practical deployment is to host the generative decoder on the shared AI-accelerated infrastructure, such as the DU/CU pool and nearby edge server, where GPU accelerators, memory, power supply, and cooling resources can be provisioned more effectively. In multi-user uplink scenarios, the generative decoder would also need to be scheduled as a shared computing resource according to service latency requirements, user priority, and accelerator availability.} \re{SUPIR is used in this work as a proof-of-concept receiver model to validate the principle of inference-capable reconstruction, rather than as a fixed implementation requirement of LiTCom. 
The current prototype is therefore more suitable for latency-tolerant or latency-flexible uplink services, while strict real-time applications would require lighter receiver models or dedicated AI inference accelerators. An evaluation of LiTCom under different GenAI decoders is left for future work.
}


		\section{{Conclusion}\label{sec:VII}}
	 	 This paper proposed LiTCom, a novel communication framework for robust 6G uplink under low SNR conditions.  LiTCom embraced the resource asymmetry between transceivers, applying lightweight coding schemes at the transmitter and a powerful GenAI model for semantic inference at the receiver. The representing sufficiency and error robustness properties of LiTCom are characterized, based on which the semantic-preserving source codes and error-distributing codes were introduced and designed, respectively. A novel semantic QoE measure was proposed to evaluate the performance of LiTCom, and the perceived coverage was defined accordingly.   Moreover,  importance-aware power allocation strategies were developed to enhance the system performance.  Simulation results validated the effectiveness of LiTCom framework, the designed lightweight coding schemes, as well as the power allocation strategies. LiTCom reduces over $95\%$  transmitter-side computations while achieving up to $8$ dB and $2.5$ dB SNR gains compared to conventional 5G and Deep-JSCC baselines. 
		 
	 	 \re{Despite these promising results, several limitations should be acknowledged. First, the current validation focuses on static natural-image uplink using an LPF-based source encoder and SUPIR as the receiver-side generative decoder. Therefore, the reported numerical gains are associated with this specific modality, encoder design, and GenAI backend. Although LiTCom is not tied to SUPIR, different generative decoders may lead to different restoration quality, degradation tolerance, inference latency, and hardware requirements. Second, the robustness of LiTCom is conditional on the residual channel-induced degradation remaining within the capability of the receiver-side GenAI model. More severe mismatch conditions, such as burst errors, high-mobility fading, imperfect CSI, or structured decoding failures, may require domain adaptation or channel-quality-aware conditioning of the generative decoder. Third, the LPF-based encoder is a deliberately simple and lightweight instantiation of semantic-preserving source coding, but it may discard high-frequency components that are task-critical for fine-grained inference or precise reconstruction.    Fourth, extending LiTCom to multi-modal sources and multi-user deployment raises additional challenges, including distributed source correlation, inter-user interference, and temporal, spatial, and semantic misalignment across modalities. Finally, the energy modelling of the GenAI models requires further investigation for holistic system-level energy evaluation.} 
\appendices

	\section{Reconstructed Images \label{appendix2}}
		
	 \begin{figure*}[tp]
	 	\centering
	 	\includegraphics[width=0.88\textwidth]{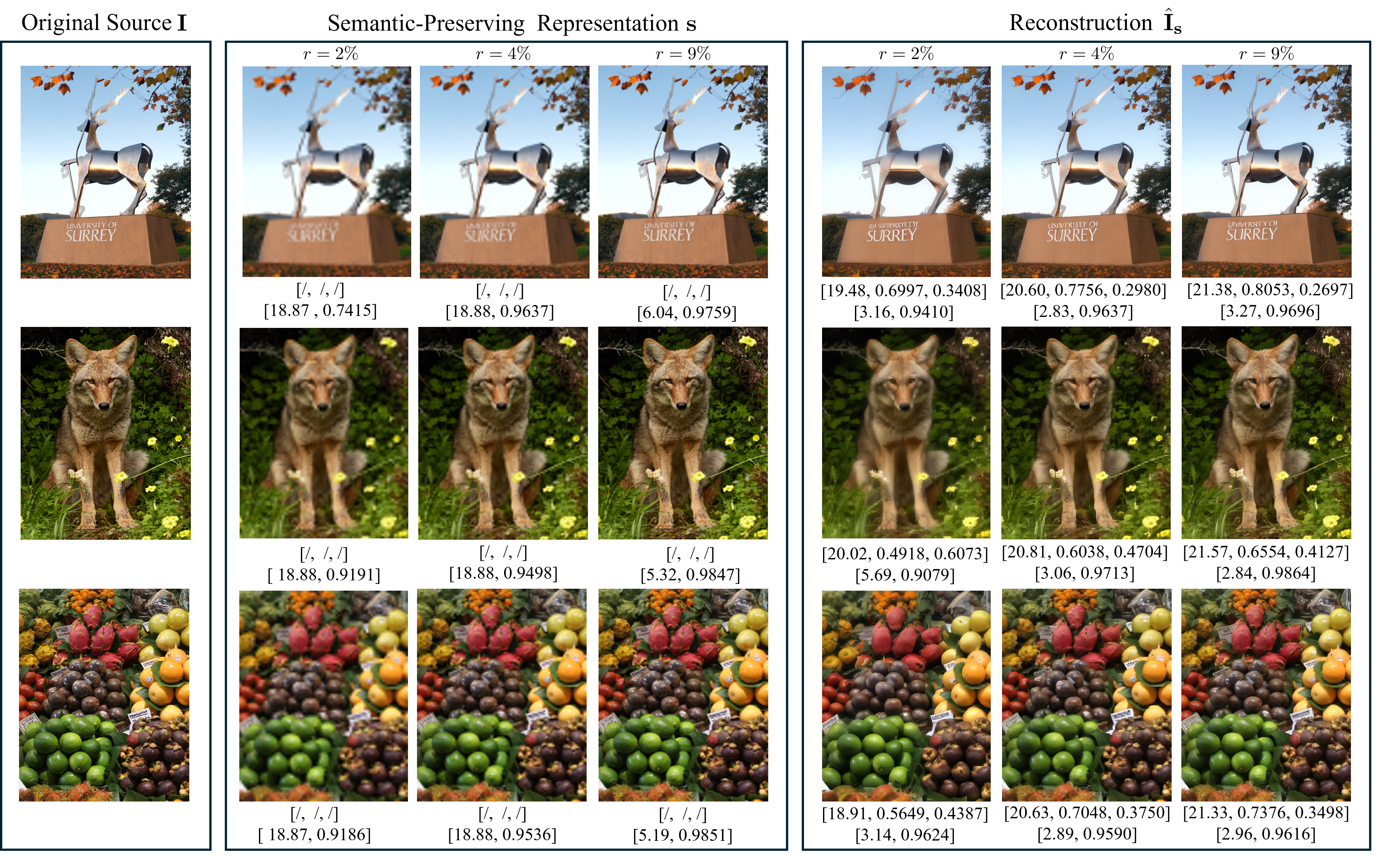}
	 	\caption{The original source, compressed representation, and the reconstructions under the proposed LiTCom framework with perfect transmission. The first row of the scores are[PSNR(dB), SSIM, LPIPS], which are not calculated for the representations due to the different image sizes. The second row of the scores are [NIQE, CLIP].}
	 	\label{fig:sourceCoding}
	 \end{figure*}
	
	\begin{figure*}[tp]
		\centering
		\includegraphics[width=0.92\textwidth]{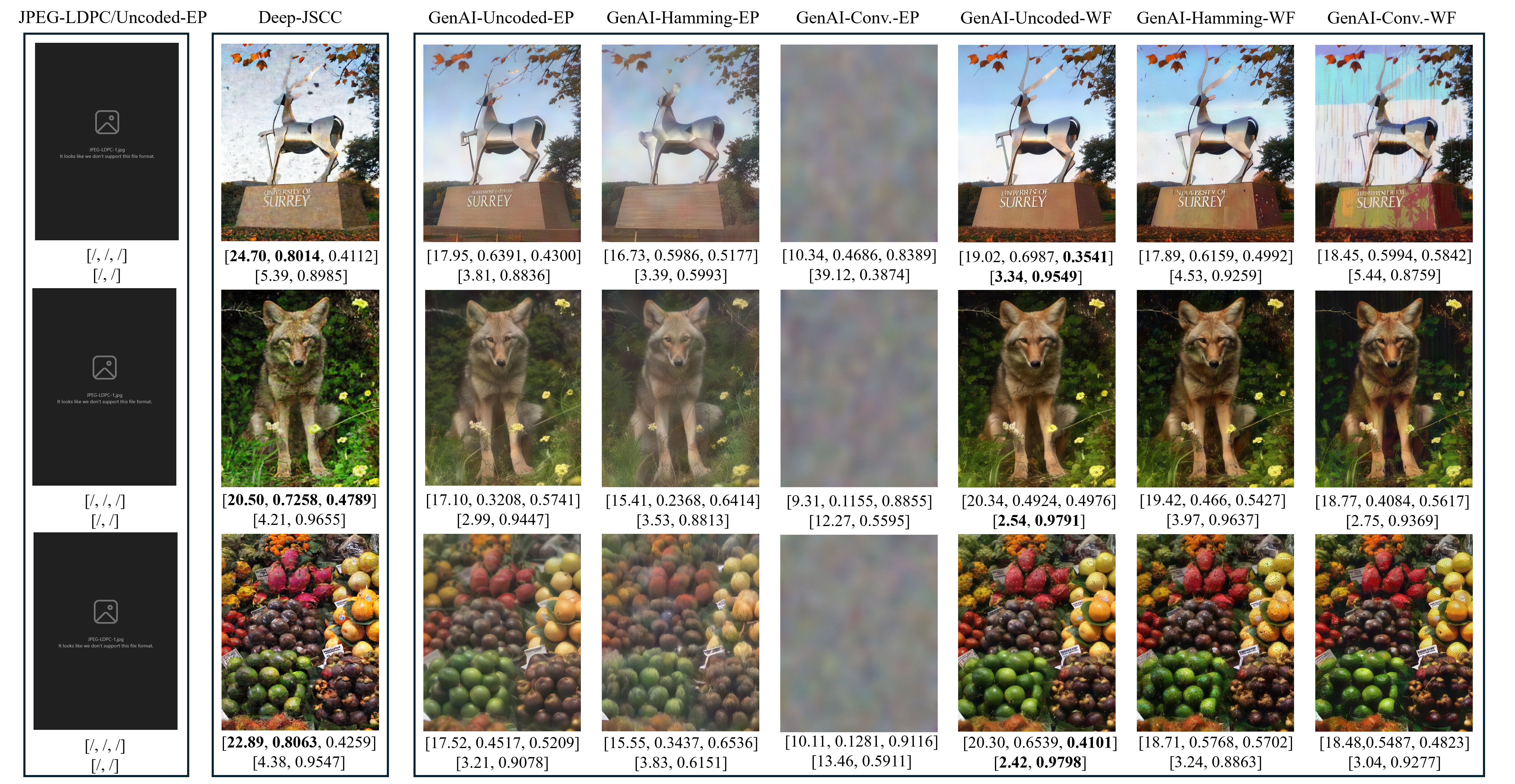}
		\caption{The reconstructions under LiTCom,   5G NR-like systems, and Deep-JSCC where $r=9\%$ and $E_b/N_o=-2$~dB. The scores are [PSNR(dB), SSIM, LPIPS] for the first row, and [NIQE, CLIP] for the second row.}
		\label{fig:channelCoding} 
	\end{figure*}
		
		\begin{figure*}[tp]
			\centering
			\includegraphics[width=0.88\textwidth]{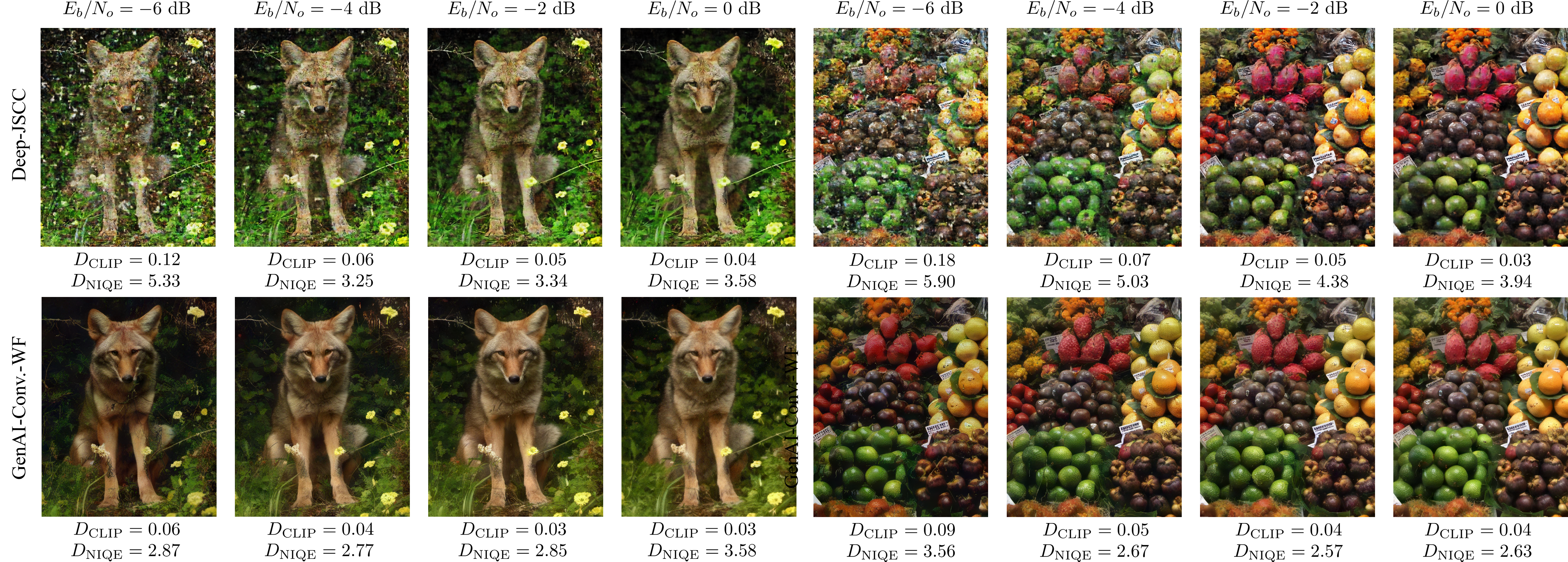}
			\caption{Visualization comparison between the proposed GenAI-Uncoded-WF and the Deep-JSCC baseline at low SNRs at a compression rate of $r=9\%$.}
			\label{fig:robustChannel}
		\end{figure*}

		\bibliographystyle{IEEEtran}
		\bibliography{reference_cleaned}

@article{giordani2020toward,
	title={Toward {6G} networks: Use cases and technologies},
	author={Giordani, Marco and Polese, Michele and Mezzavilla, Marco and Rangan, Sundeep and Zorzi, Michele},
	journal={IEEE Commun. Mag.},
	volume={58},
	number={3},
	pages={55--61},
	year={2020},
	publisher={IEEE}
}

@article{wang2023road,
	title={On the road to {6G}: Visions, requirements, key technologies, and testbeds},
	author={Wang, Cheng-Xiang and You, Xiaohu and Gao, Xiqi and Zhu, Xiuming and Li, Zixin and Zhang, Chuan and Wang, Haiming and Huang, Yongming and Chen, Yunfei and Haas, Harald and others},
	journal={ IEEE Commun. Surv. Tutor.},
	volume={25},
	number={2},
	pages={905--974},
	year={2023},
	publisher={IEEE}
}

@article{yang2022semantic,
	title={Semantic communications for future internet: Fundamentals, applications, and challenges},
	author={Yang, Wanting and Du, Hongyang and Liew, Zi Qin and Lim, Wei Yang Bryan and Xiong, Zehui and Niyato, Dusit and Chi, Xuefen and Shen, Xuemin and Miao, Chunyan},
	journal={IEEE Commun. Sur. Tutor.},
	volume={25},
	number={1},
	pages={213--250},
	year={2022},
	publisher={IEEE}
}

@techreport{3gpp_ts_38_300,
	author       = {3rd Generation Partnership Project (3GPP)},
	title        = {{5G; NR}; NR and NG-RAN Overall Description (Release 19)},
	institution  = {3GPP},
	type         = {Technical Specification (TS)},
	number       = {38.300},
	year         = {2025}
}

@techreport{AIRAN2026AIonRAN,
	author       = {AI-RAN Alliance},
	title        = {{AI-on-RAN}: Enabling Monetizable Differentiated Connectivity for {AI}},
	institution  = {{AI-RAN Alliance}},
	year         = {2026},
	type         = {White Paper},
	url          = {https://ai-ran.org/documents/AI-RAN-WG3-AI-on-RAN-Whitepaper.pdf}
}

@article{wallace1992jpeg,
	title={The {JPEG} still picture compression standard},
	author={Wallace, Gregory K},
	journal={IEEE Trans. Consumer Electronics},
	volume={38},
	number={1},
	pages={xviii--xxxiv},
	year={1992},
	publisher={IEEE}
}

@article{wiegand2003overview,
	title={Overview of the H. 264/AVC video coding standard},
	author={Wiegand, Thomas and Sullivan, Gary J and Bjontegaard, Gisle and Luthra, Ajay},
	journal={IEEE Trans. Circuits Syst. Video Technol.},
	volume={13},
	number={7},
	pages={560--576},
	year={2003},
	publisher={IEEE}
}

@article{gunduz2022beyond,
	title={Beyond transmitting bits: Context, semantics, and task-oriented communications},
	author={G{\"u}nd{\"u}z, Deniz and Qin, Zhijin and Aguerri, Inaki Estella and Dhillon, Harpreet S and Yang, Zhaohui and Yener, Aylin and Wong, Kai Kit and Chae, Chan-Byoung},
	journal={IEEE J. Sel. Areas Commun.},
	volume={41},
	number={1},
	pages={5--41},
	year={2022},
	publisher={IEEE}
}

@article{hamming1950error,
	title={Error detecting and error correcting codes},
	author={Hamming, Richard W},
	journal={The Bell System Tech. J.},
	volume={29},
	number={2},
	pages={147--160},
	year={1950},
	publisher={Nokia Bell Labs}
}

@article{viterbi1971convolutional,
	title={Convolutional codes and their performance in communication systems},
	author={Viterbi, Andrew},
	journal={IEEE Trans. Commun. Tech.},
	volume={19},
	number={5},
	pages={751--772},
	year={1971},
	publisher={IEEE}
}

@inproceedings{berrou1993near,
	title={Near Shannon limit error-correcting coding and decoding: Turbo-codes. 1},
	author={Berrou, Claude and Glavieux, Alain and Thitimajshima, Punya},
	booktitle={IEEE Int. Conf. Commun.},
	volume={2},
	pages={1064--1070},
	year={1993},
	organization={IEEE}
}

@article{richardson2001design,
	title={Design of capacity-approaching irregular low-density parity-check codes},
	author={Richardson, Thomas J and Shokrollahi, Mohammad Amin and Urbanke, R{\"u}diger L},
	journal={IEEE Trans. Inf. theory},
	volume={47},
	number={2},
	pages={619--637},
	year={2001},
	publisher={IEEE}
}

@article{kountouris2021semantics,
	title={Semantics-empowered communication for networked intelligent systems},
	author={Kountouris, Marios and Pappas, Nikolaos},
	journal={IEEE Commun. Mag.},
	volume={59},
	number={6},
	pages={96--102},
	year={2021},
	publisher={IEEE}
}

@article{shao2024theory,
	title={A theory of semantic communication},
	author={Shao, Yulin and Cao, Qi and G{\"u}nd{\"u}z, Deniz},
	journal={IEEE Trans. Mobile Computing},
	volume={23},
	number={12},
	pages={12211--12228},
	year={2024},
	publisher={IEEE}
}

@article{xie2021deep,
	title={Deep learning enabled semantic communication systems},
	author={Xie, Huiqiang and Qin, Zhijin and Li, Geoffrey Ye and Juang, Biing-Hwang},
	journal={IEEE Trans. Signal Process.},
	volume={69},
	pages={2663--2675},
	year={2021},
	publisher={IEEE}
}

@article{weng2021semantic,
	title={Semantic communication systems for speech transmission},
	author={Weng, Zhenzi and Qin, Zhijin},
	journal={IEEE J. Sel. Areas  Commun.},
	volume={39},
	number={8},
	pages={2434--2444},
	year={2021},
	publisher={IEEE}
}

@article{bourtsoulatze2019deep,
	title={Deep joint source-channel coding for wireless image transmission},
	author={Bourtsoulatze, Eirina and Kurka, David Burth and G{\"u}nd{\"u}z, Deniz},
	journal={IEEE Trans. Cognitive Commun. Networking},
	volume={5},
	number={3},
	pages={567--579},
	year={2019},
	publisher={IEEE}
}

@ARTICLE{xu2023JSCC,
	author={Xu, Jialong and Tung, Tze-Yang and Ai, Bo and Chen, Wei and Sun, Yuxuan and G{\"u}nd{\"u}z, Deniz},
	journal={IEEE Commun. Mag.}, 
	title={Deep Joint Source-Channel Coding for Semantic Communications}, 
	year={2023},
	volume={61},
	number={11},
	pages={42--48},
	doi={10.1109/MCOM.004.2200819}}

@article{erdemir2023generative,
	title={Generative joint source-channel coding for semantic image transmission},
	author={Erdemir, Ecenaz and Tung, Tze-Yang and Dragotti, Pier Luigi and G{\"u}nd{\"u}z, Deniz},
	journal={IEEE J. Sel. Areas Commun.},
	year={2023},
	publisher={IEEE}
}

@article{liu2025resitok,
	title={ResiTok: A Resilient Tokenization-Enabled Framework for Ultra-Low-Rate and Robust Image Transmission},
	author={Liu, Zhenyu and Ma, Yi and Tafazolli, Rahim},
	journal={arXiv preprint arXiv:2505.01870},
	year={2025}
}

@article{liu2024ofdm,
	title={{OFDM}-based digital semantic communication with importance awareness},
	author={Liu, Chuanhong and Guo, Caili and Yang, Yang and Ni, Wanli and Quek, Tony QS},
	journal={IEEE Trans. Commun.},
	volume={72},
	number={10},
	pages={6301--6315},
	year={2024},
	publisher={IEEE}
}

@article{zou2025analog,
	title={From analog to digital semantic communications: Architectures, challenges, and future directions},
	author={Zou, Jian and Xie, Wenwu and Xiao, Jian and Liang, Yongsheng and Moualeu, Jules M and Yang, Liang},
	journal={IEEE Wireless Commun.},
	volume={32},
	number={5},
	pages={56--62},
	year={2025},
	publisher={IEEE}
}

@article{jiang2022wireless,
	title={Wireless semantic communications for video conferencing},
	author={Jiang, Peiwen and Wen, Chao-Kai and Jin, Shi and Li, Geoffrey Ye},
	journal={IEEE J. S. Areas Commun.},
	volume={41},
	number={1},
	pages={230--244},
	year={2022},
	publisher={IEEE}
}

@article{guo2024survey,
	title={A survey on semantic communication networks: Architecture, security, and privacy},
	author={Guo, Shaolong and Wang, Yuntao and Zhang, Ning and Su, Zhou and Luan, Tom H and Tian, Zhiyi and Shen, Xuemin},
	journal={IEEE Commun. Surv. Tutor.},
	volume={27},
	number={5},
	pages={2860--2894},
	year={2024},
	publisher={IEEE}
}

@article{xu2025generative,
	title={Generative semantic communications with foundation models: Perception-error analysis and semantic-aware power allocation},
	author={Xu, Chunmei and Mashhadi, Mahdi Boloursaz and Ma, Yi and Tafazolli, Rahim and Wang, Jiangzhou},
	journal={IEEE J. S. Areas  Commun.},
	year={2025},
	publisher={IEEE}
}

@article{qiao2024latency,
	title={Latency-aware generative semantic communications with pre-trained diffusion models},
	author={Qiao, Li and Mashhadi, Mahdi Boloursaz and Gao, Zhen and Foh, Chuan Heng and Xiao, Pei and Bennis, Mehdi},
	journal={IEEE Wireless Commun. Lett.},
	year={2024},
	publisher={IEEE}
}

@article{xu2025dataimportanceJ,
	title={Data-Importance-Aware Power Allocation for Adaptive Real-Time Communication in Computer Vision Applications}, 
	author={Chunmei Xu and Yi Ma and Rahim Tafazolli and Jiangzhou Wang},
	journal={IEEE J. S. Areas  Commun.},
 	volume={43},
 	number={12},
 	pages={4015--4026},
 	year={2025},
 	publisher={IEEE}
}

@ARTICLE{11320975,
  author={Lin, Xingqin and Kundu, Lopamudra and Cammerer, Sebastian and Huang, Yan and Dick, Chris and Santhosam, Charles and Gadiyar, Rajesh and Wiesmayr, Reinhard and Studer, Christoph},
  journal={IEEE Wireless Commun.}, 
  title={{AI}-Native {6G} Empowering Intelligent {RAN} With Accelerated Compute}, 
  year={2025},
  volume={32},
  number={6},
  pages={11--14},
  doi={10.1109/MWC.2025.3620147}}

@article{chen2024big,
	title={Big {AI} models for {6G} wireless networks: Opportunities, challenges, and research directions},
	author={Chen, Zirui and Zhang, Zhaoyang and Yang, Zhaohui},
	journal={IEEE Wireless Commun.},
	year={2024},
	publisher={IEEE}
}

@ARTICLE{10384630,
	author={Bariah, Lina and Zhao, Qiyang and Zou, Hang and Tian, Yu and Bader, Faouzi and Debbah, Merouane},
	journal={IEEE Commun. Mag.}, 
	title={Large Generative {AI} Models for Telecom: The Next Big Thing?}, 
	year={2024},
	volume={62},
	number={11},
	pages={84--90},
	doi={10.1109/MCOM.001.2300364}}

@inproceedings{radford2021learning,
	title={Learning transferable visual models from natural language supervision},
	author={Radford, Alec and Kim, Jong Wook and Hallacy, Chris and Ramesh, Aditya and Goh, Gabriel and Agarwal, Sandhini and Sastry, Girish and Askell, Amanda and Mishkin, Pamela and Clark, Jack and others},
	booktitle={Proc. Int. conf. machine learning (ICML)},
	pages={8748--8763},
	year={2021},
	month={Jul.},
	address={Virtual}
}

@article{brown2020language,
	title={Language models are few-shot learners},
	author={Brown, Tom and Mann, Benjamin and Ryder, Nick and Subbiah, Melanie and Kaplan, Jared D and Dhariwal, Prafulla and Neelakantan, Arvind and Shyam, Pranav and Sastry, Girish and Askell, Amanda and others},
	journal={Advances in neural information processing systems},
	volume={33},
	pages={1877--1901},
	year={2020}
}

@inproceedings{rombach2022high,
	title={High-resolution image synthesis with latent diffusion models},
	author={Rombach, Robin and Blattmann, Andreas and Lorenz, Dominik and Esser, Patrick and Ommer, Bj{\"o}rn},
	booktitle={Proc. IEEE/CVF Conf. Comput. Vis. Pattern Recognit.},
	pages={10684--10695},
	year={2022},
	month ={Jun.},
	address ={New Orleans, Louisiana}
}

@inproceedings{zhang2023adding,
	title={Adding conditional control to text-to-image diffusion models},
	author={Zhang, Lvmin and Rao, Anyi and Agrawala, Maneesh},
	booktitle={Proc. IEEE/CVF Int. Conf. Computer Vision (ICCV)},
	pages={3836--3847},
	year={2023},
	month={Oct.},
	address={Paris, France}
}

@misc{yu2024scaling,
	title={Scaling Up to Excellence: Practicing Model Scaling for Photo-Realistic Image Restoration In the Wild}, 
	author={Fanghua Yu and Jinjin Gu and Zheyuan Li and Jinfan Hu and Xiangtao Kong and Xintao Wang and Jingwen He and Yu Qiao and Chao Dong},
	year={2024},
	eprint={2401.13627},
	archivePrefix={arXiv},
	primaryClass={cs.CV}
}

@inproceedings{lin2024diffbir,
	title={Diffbir: Toward blind image restoration with generative diffusion prior},
	author={Lin, Xinqi and He, Jingwen and Chen, Ziyan and Lyu, Zhaoyang and Dai, Bo and Yu, Fanghua and Qiao, Yu and Ouyang, Wanli and Dong, Chao},
	booktitle={Proc. Eur. Conf. Comput. Vis.},
	pages={430--448},
	year={2024},
	organization={Springer}
}

@inproceedings{yeh2017semantic,
	title={Semantic image inpainting with deep generative models},
	author={Yeh, Raymond A and Chen, Chen and Yian Lim, Teck and Schwing, Alexander G and Hasegawa-Johnson, Mark and Do, Minh N},
	booktitle={Proc. IEEE Conf. Comput. Vis.  Pattern Recog.},
	pages={5485--5493},
	year={2017}
}

@inproceedings{yu2018generative,
	title={Generative image inpainting with contextual attention},
	author={Yu, Jiahui and Lin, Zhe and Yang, Jimei and Shen, Xiaohui and Lu, Xin and Huang, Thomas S},
	booktitle={Proc. IEEE Conf. Comput. Vis.  Pattern Recog.},
	pages={5505--5514},
	year={2018}
}

@InProceedings{QoE2010,
	author={Kuipers, Fernando and Kooij, Robert
	and De Vleeschauwer, Danny
	and Brunnstr{\"o}m, Kjell"},
	title= {Techniques for Measuring Quality of Experience},
	booktitle= {Wired/Wireless Internet Communications},
	year= {2010},
	publisher = {Springer Berlin Heidelberg},
	address= {Berlin, Heidelberg},
	pages={216--227},
}

@article{zhang2018towards,
	title={Towards a {QoE} model to evaluate holographic augmented reality devices},
	author={Zhang, Longyu and Dong, Haiwei and El Saddik, Abdulmotaleb},
	journal={IEEE MultiMedia},
	volume={26},
	number={2},
	pages={21--32},
	year={2018},
	publisher={IEEE}
}

@article{zhang2023qoe1,
	title={{QoE}-driven data communication framework for consumer electronics in tele-healthcare system},
	author={Zhang, Tongguang and Zhou, Xiaokang and Liu, Jialei and Cheng, Bo and Xu, Xiaolong and Qi, Lianyong and Tian, Qiaomei and Wan, Zhiguo},
	journal={IEEE Trans. Consumer Electronics},
	volume={69},
	number={4},
	pages={719--733},
	year={2023},
	publisher={IEEE}
}

@article{zhao2016qoe,
	title={{QoE} in video transmission: A user experience-driven strategy},
	author={Zhao, Tiesong and Liu, Qian and Chen, Chang Wen},
	journal={IEEE Commun.  Surv. Tutor.},
	volume={19},
	number={1},
	pages={285--302},
	year={2016},
	publisher={IEEE}
}

@article{zhang2023qoe2,
	title={{QoE}-oriented mobile virtual reality game in distributed edge networks},
	author={Zhang, Yuan and Pu, Lingjun and Lin, Tao and Yan, Jinyao},
	journal={IEEE Trans. Multimedia},
	volume={25},
	pages={9132--9146},
	year={2023},
	publisher={IEEE}
}

@article{alreshoodi2013survey,
  title={Survey on {QoE}$\backslash${QoS} correlation models for multimedia services},
  author={Alreshoodi, Mohammed and Woods, John},
  journal={Int. J. Distrib. Parallel Syst.},
 volume={4},
 number={3},
 pages={53},
 year={2013}
}

@INPROCEEDINGS{7998249,
	author={Tahir, Bashar and Schwarz, Stefan and Rupp, Markus},
	booktitle={Int. Conf. Telecommunications (ICT)}, 
	title={BER comparison between Convolutional, Turbo, LDPC, and Polar codes}, 
	year={2017},
	volume={},
	number={},
	pages={1--7},
	doi={10.1109/ICT.2017.7998249}}

@article{gastpar2003code,
  title={To code, or not to code: Lossy source-channel communication revisited},
  author={Gastpar, Michael and Rimoldi, Bixio and Vetterli, Martin},
  journal={IEEE Trans. Inf. Theory},
  volume={49},
  number={5},
  pages={1147--1158},
  year={2003},
  publisher={IEEE}
}

@article{moliner2024blind,
	title={Blind audio bandwidth extension: A diffusion-based zero-shot approach},
	author={Moliner, Eloi and Elvander, Filip and V{\"a}lim{\"a}ki, Vesa},
	journal={IEEE/ACM Trans. Audio, Speech, Language Process.},
	volume={32},
	pages={5092--5105},
	year={2024},
	publisher={IEEE}
}

@inproceedings{li2025diffvsr,
	title={Diffvsr: Revealing an effective recipe for taming robust video super-resolution against complex degradations},
	author={Li, Xiaohui and Liu, Yihao and Cao, Shuo and Chen, Ziyan and Zhuang, Shaobin and Chen, Xiangyu and He, Yinan and Wang, Yi and Qiao, Yu},
	booktitle={Proc. IEEE Int. Conf. Computer Vision (ICCV)},
	pages={15319--15328},
	year={2025},
	month={Oct.},
	address = {Hawaii, US}
}

@inproceedings{yang2025structured,
	title={Structured {IB}: Improving information bottleneck with structured feature learning},
	author={Yang, Hanzhe and Wu, Youlong and Wen, Dingzhu and Zhou, Yong and Shi, Yuanming},
	booktitle={Proc. AAAI Conf. Artif. Intell},
	volume={39},
	number={20},
	pages={21922--21928},
	year={2025}
}

@book{goldsmith2005wireless,
	title={Wireless Communications},
	author={Goldsmith, Andrea},
	year={2005},
	publisher={Cambridge University Press}
}

@article{mittal2012making,
	title={Making a ``completely blind'' image quality analyzer},
	author={Mittal, Anish and Soundararajan, Rajiv and Bovik, Alan C},
	journal={IEEE Signal Process. Lett.},
	volume={20},
	number={3},
	pages={209--212},
	year={2012},
	publisher={IEEE}
}

@misc{xu2024semantic,
	title={Semantic-Aware Power Allocation for Generative Semantic Communications with Foundation Models}, 
	author={Chunmei Xu and Mahdi Boloursaz Mashhadi and Yi Ma and Rahim Tafazolli},
	year={2024},
	eprint={2407.03050},
	archivePrefix={arXiv},
	primaryClass={eess.SP},
	url={https://arxiv.org/abs/2407.03050}, 
}

@techreport{3gpp_ts_38_214,
	author       = {3rd Generation Partnership Project (3GPP)},
	title        = {{NR}; Physical layer procedures for data (Release 17)},
	institution  = {3GPP},
	type         = {Technical Specification (TS)},
	number       = {38.214},
	year         = {2025},
	url          = {https://www.3gpp.org/ftp/Specs/archive/38\_series/38.214}
}

@book{boyd2004convex,
	title={Convex Optimization},
	author={Boyd, Stephen and Vandenberghe, Lieven},
	year={2004},
	publisher={Cambridge University Press}
}

@book{simon2004digital,
  title={Digital communication over fading channels},
  author={Simon, Marvin K and Alouini, Mohamed-Slim},
  year={2004},
  publisher={John Wiley \& Sons}
}

@techreport{3gpp_ts_38212,
	title        = {{NR}; Multiplexing and Channel Coding},
	author       = {{3GPP}},
	institution  = {3rd Generation Partnership Project (3GPP)},
	type         = {Technical Specification (TS)},
	number       = {38.212},
	year         = {2026},
	note         = {Version 19.3.0, Release 19},
	url          = {https://www.3gpp.org/ftp/Specs/archive/38_series/38.212/}
}
		\begin{IEEEbiography}[{\includegraphics[width=1in,height=1.25in,clip,keepaspectratio]{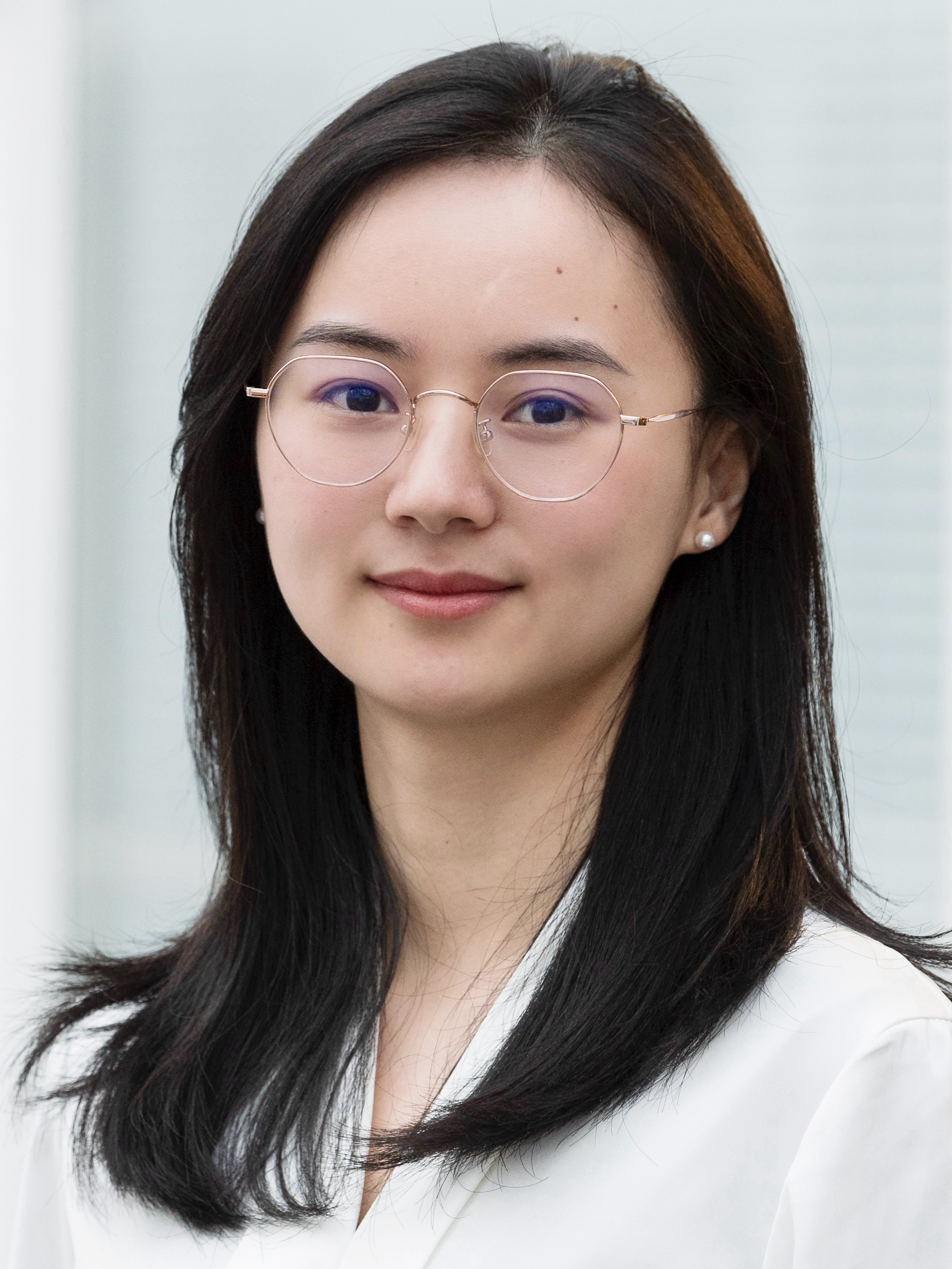}}]{Chunmei Xu} (Member, IEEE) received the B.Eng. and Ph.D. degrees from the
			School of Information Science and Engineering, Southeast University, Nanjing, China, in 2017 and 2023, respectively. She is currently a Research Fellow at the 5GIC \& 6GIC, Institute for Communication Systems, University of Surrey, Guildford, U.K. Her research focuses on  semantic communications, integrated sensing and communications and cell-free MIMO. \end{IEEEbiography}

			\begin{IEEEbiography}[{\includegraphics[width=1in,height=1.25in,clip,keepaspectratio]{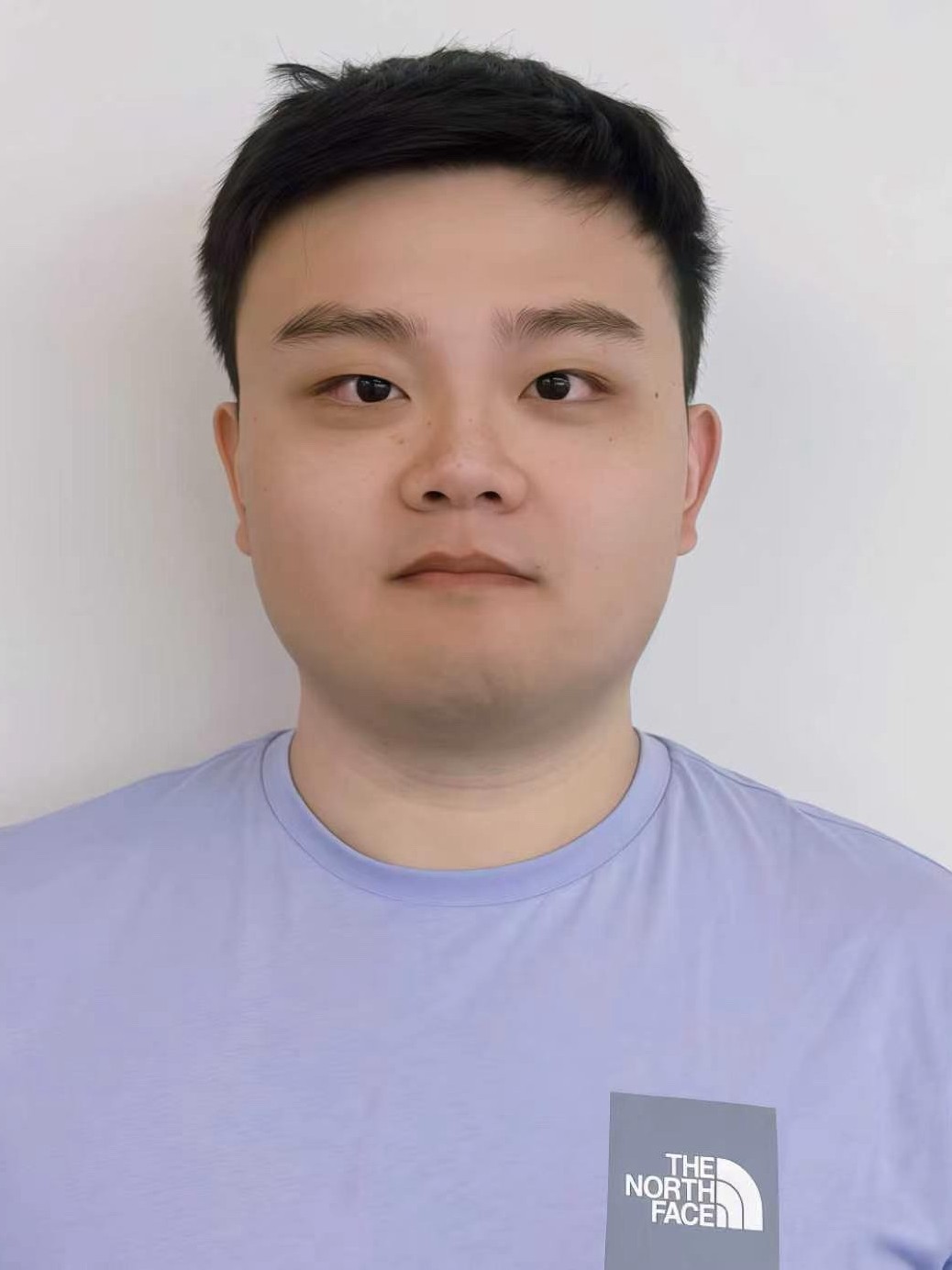}}]{Siqi Zhang} received the B.Sc. degree in electronic information technology from the Macau University of Science and Technology, Macau, China, in 2019, and the M.Sc. and Ph.D. degrees in wireless communication from the University of Surrey, U.K., in 2020 and 2025, respectively. His research interests include edge computing, semantic communication, goal-oriented communication, and communication and control co-design.
		   \end{IEEEbiography}
		
			\begin{IEEEbiography}[{\includegraphics[width=1in,height=1.25in,clip,keepaspectratio]{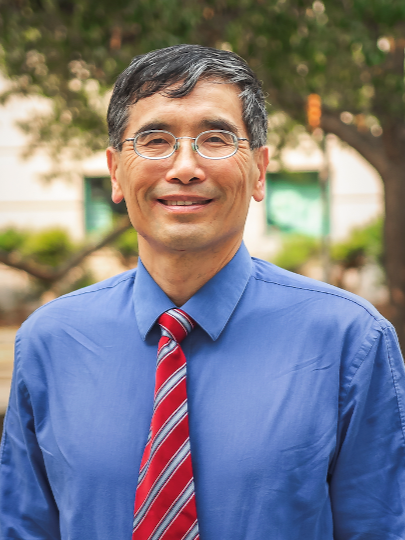}}]{Zhi Ding} 
			(Fellow, IEEE) Zhi Ding (S'88-M'90-SM'95-F'03) is with the Department of Electrical and Computer Engineering at the University of California, Davis, where he holds the position of distinguished professor. He received his Ph.D. degree in Electrical Engineering from Cornell University in 1990. From 1990 to 2000, he was a faculty member of Auburn University and later, University of Iowa. Prof. Ding joined the College of Engineering at UC Davis in 2000. His major research interests and expertise cover the areas of wireless networking, communications, signal processing, multimedia, and learning. Prof. Ding supervised 40 PhD dissertations since joining UC Davis. His research team of enthusiastic researchers works very closely with industry to solve practical problems and contributes to technological advances in signal processing, wireless networking, and learning. 
			
			Prof. Ding is a Fellow of IEEE and has served as the Chief Information Officer, Chief Marketing Officer, and Parliamentarian of the IEEE Communications Society. He was associate editor for IEEE Transactions on Signal Processing from 1994-1997, 2001-2004, and associate editor of IEEE Signal Processing Letters 2002-2005. He was a member of technical committee on Statistical Signal and Array Processing and member of technical committee on Signal Processing for Communications (1994-2003).  Dr. Ding was the General Chair of the 2016 IEEE International Conference on Acoustics, Speech, and Signal Processing and the Technical Program Chair of the 2006 IEEE Globecom. He was also an IEEE Distinguished Lecturer (Circuits and Systems Society, 2004-06, Communications Society, 2008-09). He served on as IEEE Transactions on Wireless Communications Steering Committee Member (2007-2009) and its Chair (2009-2010). Dr. Ding is a coauthor of the textbook: Modern Digital and Analog Communication Systems, 5th edition, Oxford University Press, 2019. Prof. Ding received the IEEE Communication Society?s WTC Award in 2012 and the IEEE Communication Society?s Education Award in 2020. 
			
		\end{IEEEbiography}
		
		\begin{IEEEbiography}[{\includegraphics[width=1in,height=1.25in,clip,keepaspectratio]{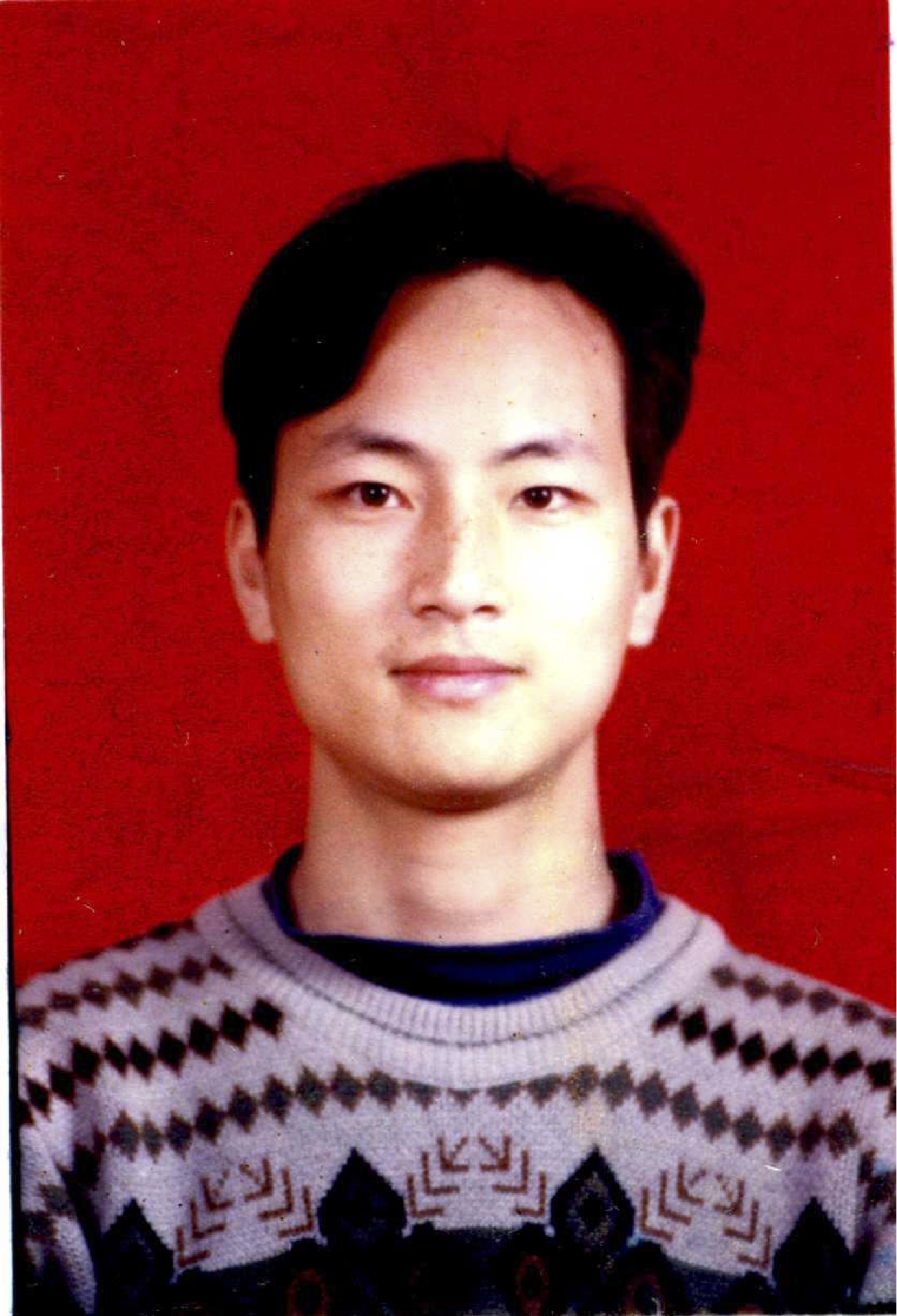}}]{Yi Ma} 
			(Senior Member, IEEE) joined the Institute for Communication Systems (ICS), University of Surrey, Guildford, U.K., in 2004, where he is currently a Chair Professor. He is the Head of Artificial Intelligence for Wireless Communication Group within the ICS to conduct the fundamental research of wireless communication systems covering signal processing, applied information theory and artificial intelligence. He is also the Chair of Air-Interface Club, ICS, and the Work Area Leader of the 5G and 6G Innovation Centre, University of Surrey. He has authored or co-authored over 150 peer-reviewed IEEE journals and conference papers in the areas of deep learning, cooperative communications, cognitive radios, interference utilization, cooperative localization, radio resource allocation, multiple-input multiple-output, estimation, synchronization, and modulation and detection techniques. He holds four international patents in the areas of spectrum sensing and signal modulation and detection. He has served as the Tutorial Chair for EuroWireless2013, PIMRC2014, and CAMAD2015. He was the Founder of the Crowd-Net Workshop in conjunction with ICC'15, ICC'16, and ICC'17. He is the Co-Chair of the Signal Processing for Communications Symposium in ICC'19.
		\end{IEEEbiography}

		 \begin{IEEEbiography}[{\includegraphics[width=1in,height=1.25in,clip,keepaspectratio]{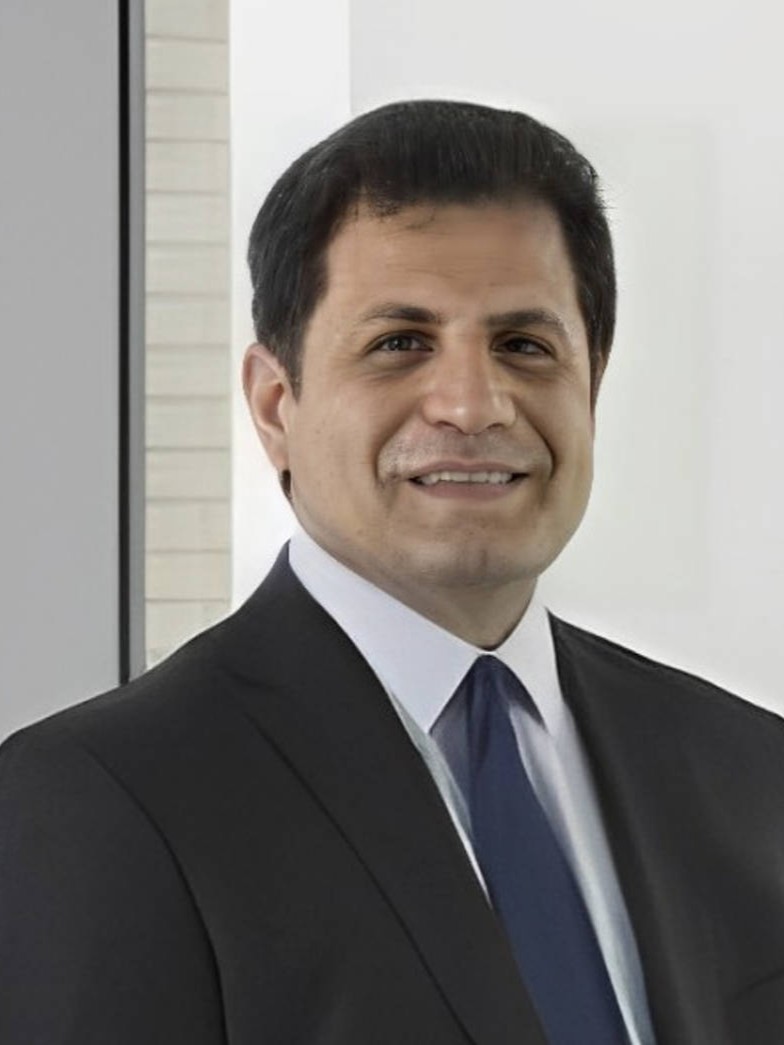}}]{Rahim Tafazolli} (Fellow, IEEE) is Regius Professor of Electronic Engineering, Professor of Mobile and Satellite Communications, Founder and Director of 5GIC, 6GIC and ICS (Institute for Communication System) at the University of Surrey. He has over 30 years of experience in digital communications research and teaching. He has authored and co-authored more than 1000 research publications and is regularly invited to deliver keynote talks and distinguished lectures to international conferences and workshops. He was the leader of study on ``grand challenges in IoT'' (Internet of Things) in the UK, 2011--2012, for RCUK (Research Council UK) and the UK TSB (Technology Strategy Board). He is the Editor of two books on Technologies for Wireless Future (Wiley) vol. 1, in 2004 and vol. 2, in 2006. He holds Fellowship of Royal Academy of Engineering, Institute of Engineering and Technology (IET) as well as that of Wireless World Research Forum. He was also awarded the 28th KIA Laureate Award- 2015 for his contribution to communications technology.
		\end{IEEEbiography}

	\end{document}